\documentclass[12pt,a4paper,epsfig]{article}
\usepackage{indentfirst}
\usepackage{a4wide}
\usepackage{epsfig}
\usepackage{amssymb}

\def\thebiblio#1{
\begin{center}\bf \large References
\end{center}
\list
{[\arabic{enumi}]}{\settowidth\labelwidth{#1.}\leftmargin\labelwidth
 \advance\leftmargin\labelsep
 \usecounter{enumi}}
 \def\newblock{\hskip .11em plus .33em minus -.07em}
 \sloppy
 \sfcode`\.=1000\relax}

\title{\textbf{Kink manifolds in a three-component scalar field theory}}
\author{A. Alonso Izquierdo$^{1,2}$, J.C. Bueno S\'anchez$^2$, M.A. Gonz\'alez
Le\'on$^2$, \\M. de la Torre Mayado$^3$  \\ \\
{\small $^1$ {\sl Department of Applied Mathematics and
Theoretical Physics.}} \\ {\small {\sl University of Cambridge.
United Kingdom.}}\\ {\small $^2$ {\sl Departamento de Matem\'atica
Aplicada. Universidad de Salamanca. Espa\~na.}} \\ {\small $^3$
{\sl Departamento de F\'{\i}sica. Universidad de Salamanca.
Espa\~na.}} }
\date{}

\begin{document}

\maketitle

\begin{abstract}
In this work we identify the manifold of solitary waves arising in
a three-component scalar field model using the Bogomol'nyi
arrangement of the energy functional. A rich structure of
topological and non-topological kinks exists in the different
sub-models contained in the theory.
\end{abstract}

\section{Introduction}

The search for solitary waves is an ongoing topic in both
Mathematics and Physics because this kind of quasi-soliton plays
an important r$\hat{{\rm o}}$le in a huge number of branches of
non-linear science. In Field Theory, they usually appear in models
that support spontaneous symmetry breaking, the most prominent
examples being kinks/domain walls, vortices, and monopoles
\cite{Rajaraman}. Starting with theories that involve a high
number of fields, the usual procedure followed to investigate the
existence of solitary waves -topological defects- is to obtain an
effective scalar field theory, imposing severe restrictions on
the original theory. In most cases, one is compelled to pursue an
effective theory that will correspond to a single scalar field
model, where the existence of topological defects can be checked
easily. The reason for this is the good-understanding of this
kind of system; as paradigmatic examples, the well-known kink and
soliton in the one-dimensional $\phi^4$ and sine-Gordon models
should be noted. Both kinds of solitary wall can be thought of as
thick walls, the topological defects in a three-dimensional
perspective. Nevertheless, the general framework is that the
effective theory depends on several scalar fields and thus the
truncation may involve an important loss of information
concerning the presence of topological defects and the structure
of spontaneous symmetry breaking. It is therefore desirable to
investigate the general properties of domain wall solutions in a
multi-scalar field theory.

In (1+1)-dimensional field theory, solitary waves are non-singular
solutions of the non-linear coupled field equations of finite
energy such that their energy density has a spacetime dependence
of the form: $\varepsilon (x,t)=\varepsilon (x-vt)$, where $v$ is
the velocity of propagation. In relativistic theories, Lorentz (or
Galilean) invariance provides all the kink solutions from the
purely static ones. The search for finite-energy static solutions
in one-dimensional field theories is tantamount to the search for
finite action trajectories in a natural dynamical system where,
the $x$-coordinate plays the r$\hat{{\rm o}}$le of time; the field
components transmute to positions in the configuration space, and
the field theoretical density energy becomes minus the mechanical
potential energy. No wonder the difficulties involved in finding
kinks in multi-component scalar field theories: one faces
multi-dimensional mechanical systems where integrability is not
ensured.

At a very early stage in the (pre-)history of the subject, a
(1+1)-dimensional field theoretical model with two real scalar
fields became relevant. Montonen and Sarker-Trullinger-Bishop
proposed the deformation of the $O(2)$-linear sigma model with a
potential energy density of $U(\phi_1,\phi_2)={1\over
2}[(\phi_1^2+\phi_2^2-1)^2+\sigma^2\phi_2^2]$, see
\cite{Montonen}. It was clear that the zeroes of the potential are
two points and hence the hunt for kinks started
immediately\footnote{We shall refer to the zeroes of the
potential as vacua throughout the paper, anticipating their
r$\hat{{\rm o}}$le in the quantization of this classical field
theory. Also, because these two points are related by the
internal symmetry group ${\Bbb Z}_2\otimes{\Bbb Z}_2$ generated
by $\phi_1\rightarrow -\phi_1$ and $\phi_2\rightarrow -\phi_2$,
we shall sometimes refer to this set as the vacuum orbit.}. Using
a trial orbit method in the associated two-dimensional mechanical
system, Rajaraman identified two different topological kinks
joining the two vacua of the system that live on a straight line
and half-ellipses respectively. Only one component of the scalar
field is non-zero in the first case, but the two-components differ
from zero in the second kind of solution; for this reason, these
solitary waves are referred to as TK1 (straight line) and TK2/TK2*
(upper/lower half-ellipse) kinks in the literature that appeared
later. Rajaraman also found one kink associated to a closed
trajectory starting from and ending at the same point of the the
vacuum orbit. Magyari and Thomas \cite{MT} realized that the
mechanical system associated with the MSTB model is integrable
-there is a second invariant in involution with the mechanical
energy- and used this fact to show that there exists a whole
family of two-component non-topological kinks (NTK2), all of them
degenerated in energy with Rajaraman's NTK2 kink; explicit kink
form factors were only described by numerical methods.

The main breakthrough in analytically finding all the solitary
waves of the MSTB model emerged in Reference \cite{Ito}. Ito
discovered that the mechanical problem was not only integrable but
that it was Hamilton-Jacobi separable by using elliptic
coordinates. In this setting, he showed the analytic formulas for
the kink orbits and the kink form factors, unveiling the
mathematical reasons for the previously observed striking kink sum
rule. Immediately, the stability of this degenerate kink family
was questioned; application of the Morse index theorem solved
this problem in \cite{ItoTasaki}. A parallel with the Morse
theory of geodesics was established somewhat later in Reference
\cite{JMG}. Thus, a clear connection arose between solitary waves,
their stability, and dynamical systems. In Reference \cite{AMJ1},
several of us showed that the MSTB model is not unique in this
respect; two (1+1)-dimensional field theoretical models with two
real scalar fields -referred to as model A and B in that paper-
have manifolds of solitary waves with similar structures. To find
the analytic expression for the kinks of model A, we were
prompted to solve an integrable dynamical system classified as
Liouville Type I, see \cite{Per}. The system belongs to the same
class as that found in the MSTB model -the two-dimensional
Garnier system \cite{Gar}- but there are three differences: (a)
the potential energy density is a polynomial of sixth order in
the fields (instead of fourth); (b) the vacuum orbit has five
points (instead of 2), and (c) there are many more stable kinks
than in the MSTB model. Model B is characterized by a fourth-order
potential energy density in the two scalar fields. The main
feature, however, is the need to solve a Liouville integrable
system of Type III, i.e. Hamilton-Jacobi separable in parabolic
coordinates. The vacuum orbit has four points and there are
manifolds of stable and unstable kinks.

In recent years, all this work has proved to be fruitful in the
framework of supersymmetric theories. In the dimensional reduction
of a generalized Wess-Zumino model with two chiral superfields,
Bazeia-Nascimento-Ribeiro-Toledo (henceforth referred to as the
BNRT model) \cite{Bazeia} found one one-component topological
kink (TK1) and one two-component topological kink (TK2). In this
case, the vacuum orbit has four points and the potential energy
density is a polynomial in the fields of order four.
Understanding the BNRT model as a deformation of model B, some of
us discovered the whole manifold of kink orbits \cite{AMJ2}. There
is kink degeneracy, also found slightly earlier by Shifman and
Voloshin in one of the topological sectors \cite{SV}, and, for
two critical values of the coupling constant, analytic formulas
for the kink form factors are available. One of them corresponds
exactly to model B; the other one leads to a Liouville system of
Type IV, Hamilton-Jacobi separable in Cartesian coordinates.
Interesting consequences have been translated to the dynamics of
intersecting branes \cite{AMJM}. How thick walls grow from
one-component kinks is well known. Composite kinks give rise to a
non-trivial low energy dynamics for intersecting walls as
geodesic motion in the kink moduli space (the space of the
integration constants with a metric inherited from the field
theoretic kinetic energy). Another supersymmetric model that
shows a rich pattern of kink solutions is the Wess-Zumino model
itself. The BPS kink states of this ${\cal N}=2$ supersymmetric
$(1+1)D$ model with a complex scalar field and holomorphic
superpotential were discovered by Vafa et al. in \cite{Va}. In
\cite{AMJ3}, two of us studied this system from the point of view
of the real-analytic structure. The vacuum orbit having been
identified, the flow between the vacuum points was determined as
the gradient of the real(imaginary) part of the superpotential.
Thus, kink orbits are identified with real algebraic curves.

Here, we continue to struggle with the extension of these studies
to field theoretical models with three real scalar fields. In
\cite{Nos3}, some of us explored the generalization of the MSTB
model. The solution of the three-dimensional Garnier system using
three-dimensional Jacobi coordinates revealed the existence of an
extremely complex variety of kinks. Nevertheless, the structure
of the kink manifold and its stability was completely unraveled
in \cite{Nos4}. The main goal of the present paper is to identify
the kink manifold arising in a family of three-component
relativistic field models with a vacuum manifold that contains
several elements or points. This family can be interpreted as the
natural generalization of the generalized MSTB model studied in
\cite{Nos3}\cite{Nos4} in the sense of St\"ackel-type systems. The
most interesting feature of this generalization is that the
number of elements in the vacuum manifold depends on the range of
relative values of the coupling constants. Therefore, we can find
different submodels of our system, which have a very rich
structure of kink manifolds. When the energy density of these
kinks or solitary waves is studied we find that several families
of these solutions are degenerate, which allows us to claim that
some kink families indeed consist of more basic kinks, such that
their energy density displays several lumps associated with the
basic kinks. In our model, we are able to find solutions with
two, three or four lumps.

The organization of the paper is as follows: In Section 2 we
introduce the model, writing the expressions in St\"ackel form
and describing the different spontaneous symme\-try-brea\-king
scenarios. Section 3 is divided into four sub-sections. In 3.1, we
identify first-order differential equations satisfied by the kink
solutions, reproducing the Bogomol'nyi procedure in this context.
Sub-section 3.2 contains the resolution of these equations. In
3.3, we determine the regions where the solutions live and,
finally, in Sub-section 3.4, some general comments about the
determination of the stability of the kink solutions are offered.
In Section 4 we describe the behaviour of solitary wave families
in one of the regimes of the model, at the same time discussing
their stability properties. Finally, in Section 5 we address some
points concerning the different extensions of the model.

\section{The model}

We focus our attention on the search for kink solutions arising in
three-component scalar field models in a (1+1) Minkowskian
space-time, whose dynamics is governed by the action functional
\[
S[\phi]=\int d^2 x \left[ \frac{1}{2} \sum_{j=1}^3 \partial_\mu
\phi_j \partial^\mu \phi_j-U(\phi)  \right]\quad ,
\]
where we use Einstein's convention for Greek indices with the
usual metric $\eta_{11}=-\eta_{22}=1$, $\eta_{12}=\eta_{21}=0$,
and where $U(\phi)$ is a smooth non-negative function that
depends on the three-component scalar field
$\phi=(\phi_1,\phi_2,\phi_3)$. We use natural units, hence $c=1$,
and we shall henceforth denote $x^0\equiv t$ and $x^1\equiv x$.
The Euler-Lagrange equations in this case are written as the
following system of second-order partial differential equations
\begin{equation}
\frac{\partial^2 \phi_i}{\partial t^2}-\frac{\partial^2
\phi_i}{\partial x^2}=-\frac{\partial U}{\partial \phi_i}
(\phi_1,\phi_2,\phi_3) \hspace{0.5cm} i=1,2,3\, . \label{eq:e1}
\end{equation}

Kinks are finite-energy solutions of (\ref{eq:e1}), such that the
time dependence is dictated by the Lorentz invariance:
$\phi_K(t,x)=\phi(\frac{x-vt}{\sqrt{1-v^2 }})$, and they can be
interpreted as extremals of the positive semi-definite energy
functional
\begin{equation}
{\cal E}[\phi]=\int d x \, \varepsilon(x)=\int d x \left\{
\frac{1}{2} \sum_{i=1}^3 \frac{\partial\phi_i}{\partial x}
\frac{\partial \phi_i}{\partial
x}+U(\phi_1,\phi_2,\phi_3)\right\}\quad ,\label{energy}
\end{equation}
which maintains this functional finite: ${\cal E}[\phi]<+\infty$,
see \cite{Rajaraman}. Therefore, solitary waves must comply with
the asymptotic conditions
\begin{equation}
a) \hspace{0.5cm} \lim_{x\to \pm \infty} \phi \in {\cal M}
\hspace{2cm} b) \hspace{0.5cm} \lim_{x\to \pm \infty}
\frac{d\phi}{dx}=0\quad , \label{eq:condi}
\end{equation}
where ${\cal M}$ is the set of zeroes or absolute minima of the
potential term -that is, ${\cal M}=\{(\phi_1,\phi_2,\phi_3)\in
{\mathbb R}^3 / U(\phi_1,\phi_2,\phi_3)=0\}$- which are usually
referred to as vacua of the theory because the elements of ${\cal
M}$ play this r$\hat{\rm o}$le in the corresponding quantum
theory.

The usual procedure for tackling the search for kinks in this
kind of theory is to interpret (\ref{energy}) as the action
functional of a mechanical system in which we think of the
variable $x$ as ``time"; $\phi$ as the coordinates of a unit-mass
point particle, and $V=-U$ as the potential function. From this
point of view, (\ref{eq:e1}) are merely equations of motion in the
new system. In reference \cite{Nos3}, the authors deal with the
model involving the potential function
\begin{equation}
U(\phi_1,\phi_2,\phi_3)=\frac{1}{2}
(\phi_1^2+\phi_2^2+\phi_3^2-1)^2+\frac{1}{2} \sigma_2^2
\phi_2^2+\frac{1}{2} \sigma_3^2 \phi_3^2\label{potentialMSTB3}
\end{equation}
and show that the mechanical analogue is not only completely
integrable but also Hamilton-Jacobi separable by using a system of
three-dimensional elliptic coordinates. In \cite{Nos4}, the
stability properties of kinks are analyzed and a new approach to
search for kinks based on the Bogomol'nyi decomposition are given
in the above system. The authors prove the equivalence between
the Hamilton-Jacobi equation and the Bogomol'nyi approach. The
potential function (\ref{potentialMSTB3}) has two zeroes,
$v^-=(-1,0,0), \, v^+=(1,0,0)$. Therefore, the kinks in this model
can be classified into topological and non-topological kinks
according to whether the solution connects two different vacua
(open orbits) or the solution departs and arrives at a vacuum
(closed orbits).

The search for new integrable models is not an easy task. In this
sense, we would remark the following quotation from Jacobi in his
``Vorlesungen \"uber Dynamik", which allows us to see the issue
from a different perspective:  {\it ``The main difficulty in
integrating given differential equations is to introduce suitable
variables which cannot be found by a general rule. Therefore, we
must go in the opposite direction and, after finding some
remarkable substitution, look for problems to which it could be
successfully applied".}

The goal of this paper is to generalize the above model, focusing
our attention on models with a greater-than-two number of elements
in ${\cal M}$, such that we can find a more sophisticated
symmetry-breaking scenario and a richer plethora of solitary waves
than before.

Using the same notation as in the reference \cite{Nos3}, we now
introduce a system of Jacobi elliptic coordinates ${\bf
\lambda}=(\lambda_1,\lambda_2,\lambda_3)$, with constants
$\bar\sigma_3^2=1-\sigma_3^2$, $\bar\sigma_2^2=1-\sigma_2^2$ and
$1$, which is defined as:
\begin{eqnarray}
\phi_1^2&=&\frac{1}{\sigma_2^2 \sigma_3^2}
(1-\lambda_1)(1-\lambda_2)(1-\lambda_3)\nonumber \\
\phi_2^2&=&\frac{-1}{\sigma_2^2 (\sigma_3^2-\sigma_2^2)}
(\bar\sigma_2^2-\lambda_1)(\bar\sigma_2^2-\lambda_2)(\bar\sigma_2^2-
\lambda_3) \label{eq:coor} \\
\phi_3^2&=&\frac{-1}{\sigma_3^2 (\sigma_2^2-\sigma_3^2)}
(\bar\sigma_3^2-\lambda_1)(\bar\sigma_3^2-\lambda_2)(\bar\sigma_3^2-
\lambda_3)\, ,\nonumber
\end{eqnarray}
in which the range of the coordinates is:
\begin{equation}
-\infty < \lambda_1 < \bar\sigma_3^2<\lambda_2
<\bar\sigma_2^2<\lambda_3<1\  .\label{rangos}
\end{equation}

It should be noted that this coordinate transformation is
invariant under the group $G=\mathbb{Z}_2^{\times 3}$ generated by
$\phi_a\to(-1)^{\delta_{ab}}\phi_a;\,b=1,2,3$.

Invoking (\ref{eq:coor}), the energy functional can be written as
\begin{equation}
{\cal E}[\phi]=\int d x \left\{ \frac{1}{2} \sum_{j=1}^3
g_{jj}(\lambda)
 \frac{\partial\lambda_j}{\partial x} \frac{\partial
\lambda_j}{\partial x}+U(\lambda_1,\lambda_2,\lambda_3)\right\}\
, \label{eq:e2}
\end{equation}
where the metric coefficients $g_{jj}(\lambda)=-\frac{1}{4}
\frac{f_j(\lambda)}{(\lambda_j-1)(\lambda_j-\bar\sigma_2^2)(\lambda_j-\bar\sigma_3^2)}$
have been introduced. Here, we set $\displaystyle f_j(\lambda)=
\prod_{k=1,k\neq j}^3 (\lambda_j-\lambda_k)$.

In the new variables, the potential (\ref{potentialMSTB3}) is
written as:
\begin{equation}
U(\lambda)=\frac{1}{2}
\sum_{i=1}^3\frac{\lambda_i^2(\lambda_i-\overline{\sigma}_2^2)
(\lambda_i-\overline{\sigma}_3^2)}{f_i(\lambda)}\  ,
\label{eq:garnier}
\end{equation}
and their zeroes $v^-$ y $v^+$ are mapped to one point $v\equiv
(\lambda_1^v,\lambda_2^v,\lambda_3^v)=(0,\bar{\sigma}_3^2,\bar{\sigma}_2^2)$
in the elliptic space because of the above-mentioned invariance.

In order to generalize expression (\ref{eq:garnier}), we introduce
the following potential function
\begin{equation}
U(\lambda;\bar{\alpha}^2)=\sum_{i=1}^3
U_i(\lambda;\bar{\alpha}^2) = \frac{1}{2}
\sum_{i=1}^3\frac{\lambda_i^2(\lambda_i-\overline{\sigma}_2^2)
(\lambda_i-\overline{\sigma}_3^2)(\lambda_i-\bar{\alpha}^2)^2}{f_i(\lambda)}\
, \label{eq:pot}
\end{equation}
which becomes a polynomial function of eighth degree in the
original fields. Notice that we have added a new factor
$(\lambda_i-\bar{\alpha}^2)^2$ to each of the summands in
(\ref{eq:garnier}). Thus, (\ref{eq:pot}) introduces new degenerate
vacua in ${\cal M}$, which for fixed $\overline{\sigma}_2^2$ and
$\overline{\sigma}_3^2$ depend upon the value of the coupling
constant $\bar{\alpha}^2=1-\alpha^2$. Therefore, new scenarios of
spontaneous symmetry-breaking and a richer kink manifold arise in
this model. Taking into account the range (\ref{rangos}) for the
elliptic coordinates and formula (\ref{eq:pot}), we can observe
that the new structure of the set ${\cal M}$ depends on the
relative values between the constant $\bar{\alpha}^2$ and the
fixed constants $\bar{\sigma}_2^2$, $\bar{\sigma}_3^2$ and $1$.
For instance, for $\bar{\alpha}^2>1$ the new factor
$(\lambda_i-\bar{\alpha}^2)^2$ does not vanish for any value of
$\lambda_i$ and therefore ${\cal M}$ has the same structure as
that in model (\ref{eq:garnier}). However, for $\bar{\alpha}^2\in
(\bar{\sigma}_3^2,\bar{\sigma}_2^2)$ we find new vacua located at
the points $\lambda\equiv
(\bar{\sigma}_3^2,\bar{\alpha}^2,\bar{\sigma}_2^2)$ and
$\lambda\equiv (0,\bar{\alpha}^2,\bar{\sigma}_2^2)$.

We shall now introduce different scenarios for our model depending
on the value of the constant $\bar{\alpha}^2$. We shall
distinguish the number of vacua in each case.
\begin{itemize}
\item Regime $E1$: As mentioned above, for $\bar{\alpha}^2\in \left( 1,\infty \right)$
there exists only one vacuum in the elliptic space, minimizing
the potential function:
$\lambda^{v_1}=(0,\bar{\sigma_3}^2,\bar{\sigma_2}^2)$. We have a
similar situation if the constant $\bar{\alpha}^2$ takes the
discrete values $0$, $\overline{\sigma }_3^2$ or $\overline{\sigma
}_2^2$. For this reason we define the set $L_0=\left\{
0,\overline{\sigma }_3^2, \overline{\sigma }_2^2\right\} \cup
\left( 1,\infty \right)$, taking into account that if
$\bar{\alpha}^2\in L_0$ our model only has a vacuum,
$\lambda^{v_1}$, in the elliptic space. In the Cartesian space,
the vacuum manifold ${\cal M}$ can be regarded as the orbit
generated by the action of the group $G/H_1$ over the vacuum
$v_1$, where $H_1=\mathbf{1}\times\mathbb{Z}_2\times\mathbb{Z}_2$
is the group that leaves the coordinates of $v_1$ invariant.
There are therefore two vacua in the Cartesian space ${\cal
M}_0=\{ \phi^{v^1}=(\pm 1,0,0)$\}.

The kink solutions in this model display the same behaviour as
those of the model studied in \cite{Nos3}, although the explicit
expression of the equations of motion is more complicated because
we have a polynomial of degree eight in the original fields. Owing
to this similarity, we shall not deal with this regime in our
study.

\item Regime $E2$: We now consider the range $\bar{\alpha}^2\in L_1=
\left(0,\overline{\sigma }_3^2\right)$ for the coupling constant.
In this regime, new zeroes of the potential arise on the plane
$\lambda_1=\bar{\alpha}^2$, in the elliptic space that
corresponds to the ellipsoid
$\frac{\phi_1^2}{1-\bar{\alpha}^2}+\frac{\phi_2^2}{\bar{\sigma}_2^2-\bar{\alpha}^2}
+\frac{\phi_3^2}{\bar{\sigma}_3^2-\bar{\alpha}^2}=1$ in the
Cartesian space. In fact, two vacua arise in the elliptic space,
$\lambda^{v^1}=(0,\bar{\sigma_3}^2,\bar{\sigma_2}^2)$ and
$\lambda^{v^2}=(\bar{\alpha}^2,\bar{\sigma_3}^2,\bar{\sigma_2}^2)$,
both invariant under the subgroup $H_2=H_1$. Correspondingly,
there are four vacua in the Cartesian space that correspond to the
orbit $\bigsqcup_{i=1}^2(G/H_i)v_i$. Therefore, we have ${\cal
M}_1=\{ \phi^{v_1}=(\pm 1,0,0), \phi^{v_2}=(\pm \alpha,0,0)$\}, as
depicted in Figure 1.

It is interesting to remark that the range of values
$\bar{\alpha}^2\in (-\infty, 0)$ is formally analogous to that in
which $\bar{\alpha}^2 \in L_1$, interchanging the r$\hat{\rm
o}$les of the ellipsoids $\lambda_1=0$ and
$\lambda_1=\bar{\alpha}^2$ in the previous reasoning. We shall
therefore focus our attention on the range of values $L_1$.

\begin{figure}[htbp] \hspace{4.5cm} \epsfig{file=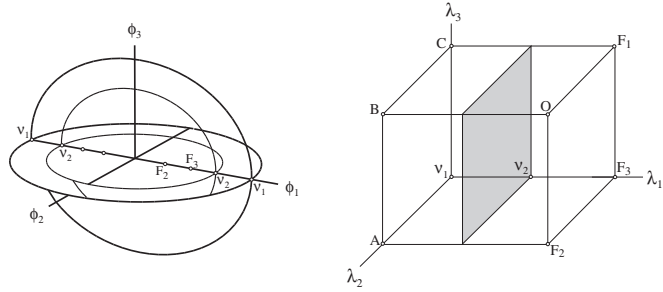,
height=4.0cm} \caption{\small \it Vacuum manifold in the Cartesian
and elliptic spaces: Regime $E2$. F$_1$, F$_2$ and F$_3$ stand
for the foci of the ellipsoid; B and C are the extremes of the
minor semi-axis, and A represents the umbilical points.}
\end{figure}

\item Regime $H1$: In this case, $\bar{\alpha}^2\in L_2=\left(
\overline{\sigma }_3^2,\overline{\sigma }_2^2\right)$. From the
values of $\bar{\alpha}^2$ and the range of the elliptic
coordinates, the zeroes of the potential term (\ref{eq:pot}) arise
on the plane $\lambda_2=\bar{\alpha}^2$, which is equivalent to
the hyperboloid of one sheet
$\frac{\phi_1^2}{1-\bar{\alpha}^2}+\frac{\phi_2^2}{\bar{\sigma}_2^2-\bar{\alpha}^2}=1+
\frac{\phi_3^2}{\bar{\alpha}^2-\bar{\sigma}_3^2}$ in the Cartesian
space. We find three vacua located at the points $\lambda^{v^1}=(
0,\bar\sigma_3^2,\bar\sigma_2^2)$,
$\lambda^{v^2}=(\bar\sigma_3^2,\bar{\alpha}^2,\bar\sigma_2^2)$ and
$\lambda^{v^3}=( 0,\bar{\alpha}^2,\bar\sigma_2^2)$. The vacua
$v_1$ and $v_2$ remain invariant under $H_1$, whereas $v_3$ is
invariant under
$H_3=\mathbf{1}\times\mathbb{Z}_2\times\mathbf{1}$. There are
eight vacua in the Cartesian space corresponding to the orbit
$\bigsqcup_{i=1}^3(G/H_i)v_i$, with coordinates ${\cal
M}_2=\{\phi^{v_1}=(\pm 1,0,0), \phi^{v_2}=(\pm
\alpha,0,0),\phi^{v_3}=(\pm \frac{\alpha}{\sigma_3},0,\pm
\frac{\bar{\sigma}_3
}{\sigma_3}\sqrt{\bar{\alpha}^2-\bar{\sigma}_3^2})\}$, as shown
in Figure 2. In section 4, for the sake of clarity we shall focus
on this regime in order to describe in detail a particular kink
manifold of the model instead of discussing it in each single
regime.

\begin{figure}[htbp] \hspace{4.5cm} \epsfig{file=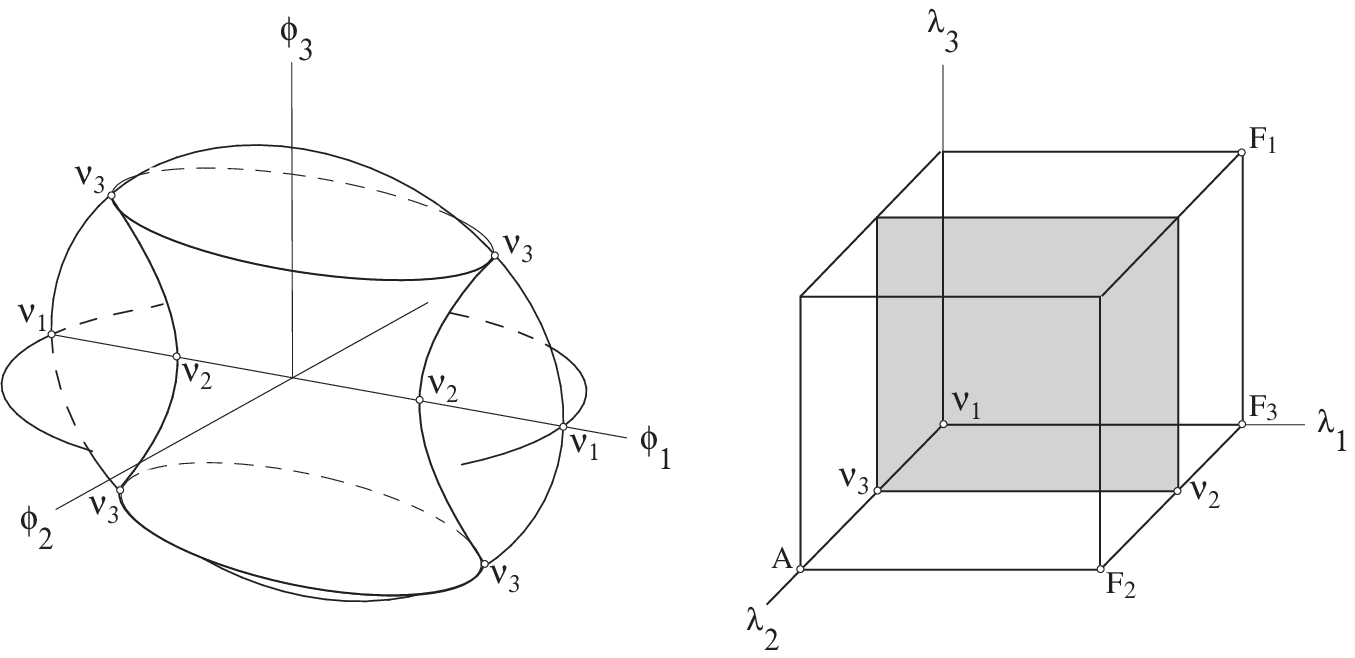,
height=4.0cm} \caption{\small \it Vacuum manifold in the Cartesian
and elliptic spaces: Regime $H1$.}
\end{figure}

\item Regime $H2$: This case is characterized by
$\bar{\alpha}^2\in L_3=\left( \overline{\sigma }_2^2,1\right)$.
Applying the same reasoning as before, we find that new vacua
arise on the plane $\lambda_3=\bar{\alpha}^2$; that is, the
hyperboloid of two sheets
$\frac{\phi_1^2}{1-\bar{\alpha}^2}=1+\frac{\phi_2^2}{\bar{\alpha}^2-\bar{\sigma}_2^2}+
\frac{\phi_3^2}{\bar{\alpha}^2-\bar{\sigma}_3^2}$ in Cartesian
coordinates. In particular, the potential has four minima:
$\lambda^{v_1}=( 0,\bar\sigma_3^2,\bar\sigma_2^2)$;
$\lambda^{v_2}=( \bar\sigma_3^2,\bar\sigma_2^2,\bar{\alpha}^2)$;
$\lambda^{v_3}=( 0,\bar\sigma_2^2,\bar{\alpha}^2)$, and
$\lambda^{v_4}=( 0,\bar\sigma_3^2,\bar{\alpha}^2)$. The vacuum
$v_4$ is invariant under
$H_4=\mathbf{1}\times\mathbf{1}\times\mathbb{Z}_2$, and the
Cartesian vacuum manifold is the orbit
$\bigsqcup_{i=1}^4(G/H_i)v_i$; namely,
\begin{eqnarray*}
{\cal M }_3&=&\left\{ \phi^{v_1}=(\pm 1,0,0),\phi^{v_2}=(\pm
\alpha,0,0),\right.\\ && \left.\phi^{v_3}=\left(\pm
\frac{\alpha}{\sigma_3},0,\pm \frac{\bar{\sigma}_3
}{\sigma_3}\sqrt{\bar{\alpha}^2-\bar{\sigma}_3^2}\right),
\phi^{v_4}=\left(\pm \frac{\alpha}{\sigma_2},\pm
\frac{\bar\sigma_2}{\sigma_2}
\sqrt{\bar{\alpha}^2-\bar\sigma_2^2},0\right)\right\}.
\end{eqnarray*}

\begin{figure}[htbp] \hspace{4.5cm} \epsfig{file=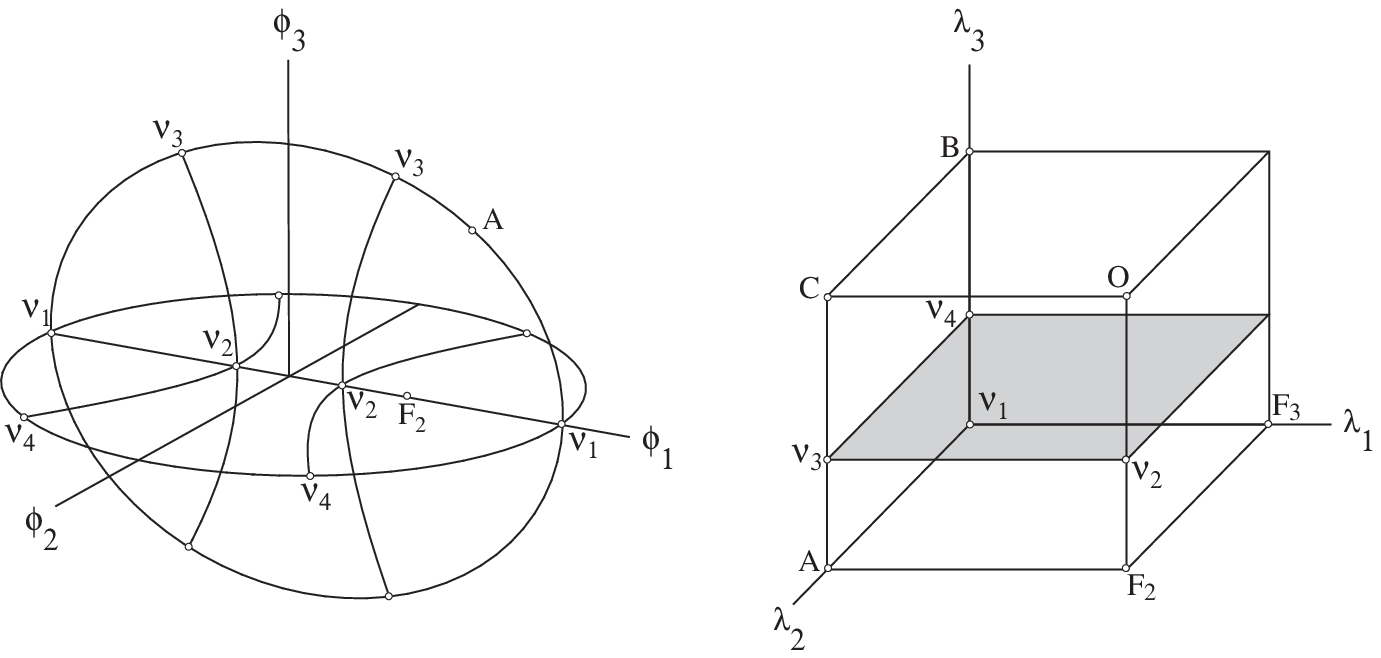,
height=4.0cm} \caption{\small \it Vacuum manifold in the Cartesian
and elliptic spaces: Regime $H2$.}
\end{figure}

\item Regime $H2\,'$:  In this case, the coupling constant
$\bar{\alpha}^2$ is set equal to unity. In the internal elliptic
space we can read the minima as $\lambda^{v_1}=(
0,\bar\sigma_3^2,\bar\sigma_2^2)$, $\lambda^{v_2}=(
\bar\sigma_3^2,\bar\sigma_2^2,1)$, $\lambda^{v_3}=(
0,\bar\sigma_2^2,1)$ and $\lambda^{v_4}=( 0,\bar\sigma_3^2,1)$,
which correspond to eight minima in the Cartesian space: ${\cal
M}_3^{\,'}=\displaystyle\lim_{\bar{\alpha}^2\to 1}{\cal
M}_3=\{\phi^{v_1}=(\pm 1,0,0), \phi^{v_2}=(0,0,0),
\phi^{v_3}=(0,0,\pm \bar\sigma_3), \phi^{v_4}=(0,\pm
\bar\sigma_2,0)\}$.

In this latter case, the plane $\lambda_3=1$ is introduced into
the elliptic space. Unlike the previously introduced planes, this
is no longer a regular one, and this can be readily seen in the
degeneracy exhibited by the $H$2 vacuum manifold at the limit
$\bar{\alpha}^2\to 1$; this singular plane corresponds to the
plane $\phi_1=0$ in Cartesian coordinates. Regarding the kink
manifold, this is basically the same as that of the $H$2 model,
except that the kink solutions existing on the two sheets of the
hyperboloid and in between them now degenerate into kink
solutions on the plane $\phi_1=0$. This situation is the 3D
analogue of model A in \cite{AMJ1}.
\end{itemize}

\section{First-order equations and Kink Manifolds}

\subsection{The superpotential and the Bogomol'nyi arrangement}

We notice that the potential (\ref{eq:pot}) determines a St\"ackel
system \cite{Per}. Therefore, the Hamilton-Jacobi equation of the
mechanical analogue is separable using the system of Jacobi
elliptic coordinates. However, here we shall make use of the
Bogomol'nyi arrangement in order to obtain the kink manifold of
our model. The two procedures are equivalent (see \cite{Nos4}) but
the second one allows us to identify the supersymmetric extension
of our field theory, given that if the energy functional
(\ref{eq:e2}) can be written as
\begin{equation}\label{eq:e3} {\cal E}[\phi]=\int d x \frac{1}{2}
\sum_{j=1}^3 g_{jj} \left( \frac{\partial\lambda_j}{\partial x}\pm
\frac{1}{g_{jj}}\frac{\partial W(\lambda)}{\partial \lambda_j}
\right)^2 \mp \int d x \frac{d W(\lambda)}{d x} \end{equation}
for some function $W(\lambda_i)$, then the underlying field
theory has a supersymmetric extension in which the function $W$
plays the r$\hat{\rm o}$le of superpotential in the
supersymmetric field theory, see \cite{NosAOP}. Therefore, the
superpotential $W$ must comply with
\begin{displaymath}
2U(\lambda)=g_{11}^{-1}(\lambda)\left(\frac{\partial
W}{\partial\lambda_1}\right)^2+g_{22}^{-1}(\lambda)\left(\frac{\partial
W}{\partial\lambda_2}\right)^2+g_{33}^{-1}(\lambda)\left(\frac{\partial
W}{\partial\lambda_3}\right)^2\  .
\end{displaymath}

Plugging the expression of the potential function (\ref{eq:pot})
and the metric coefficients into the above equation, we have
\[
\sum_{i=1}^3\frac{\lambda_i^2(\lambda_i-\bar{\alpha}^2)^2
\prod_{j=2}^3(\lambda_i-\overline{\sigma}_j^2)}
{f_i(\lambda)}=\sum_{i=1}^3
\frac{-4(\lambda_i-1)\prod_{j=2}^3(\lambda_i-\overline{\sigma}_j^2)}
{f_i(\lambda)}\left(\frac{\partial
W}{\partial\lambda_i}\right)^2\  ,
\]
which can be solved easily by the ansatz
$W=W_1(\lambda_1)+W_2(\lambda_2)+W_3(\lambda_3)$. The three
resulting decoupled ordinary differential equations
\[
\left(\frac{dW_i}{d\lambda_i}\right)^2=\frac{\lambda_i^2(\lambda_i-\bar{\alpha}^2)^2}
{4(1-\lambda_i)}\quad,\quad i=1,2,3
\]
lead us to the expression of the superpotential function $W$
\begin{displaymath}
W^{(\beta_1,\beta_2,\beta_3)}(\lambda)=\sum_{i=1}^3
W^{\beta_i}_i(\lambda_i)=\frac{1}{15}
\sum_{i=1}^3(-1)^{\beta_i}P_2(\lambda_i)\sqrt{1-\lambda_i}\
\quad,\quad\beta_i=0,1\  ,
\end{displaymath}
where $P_2(\lambda_i)=2d+d\lambda_i-3\lambda_i^2$, with
$d=(5\bar{\alpha}^2-4)$.

Extremal trajectories for the energy functional (\ref{eq:e3})
arise if the following system of first-order differential
equations
\begin{eqnarray}\label{eq:foe}
\frac{d\lambda_i}{dx}&=&(-1)^{\beta_i}g_{ii}^{-1}(\lambda)\frac{dW_i}{d\lambda_i}\nonumber\\
&=&(-1)^{\beta_i}2\frac{\lambda_i(\lambda_i-\bar{\alpha}^2)
(\lambda_i-\overline{\sigma}_2^2)(\lambda_i-\overline{\sigma}_3^2)}
{f_i(\lambda)} \sqrt{1-\lambda_i}
\end{eqnarray}
where $\beta_i=0,1$ and $i=1,2,3$ is satisfied, because the
squared terms in (\ref{eq:e3}) are always positive and the last
one is a constant. Due to the indeterminacy of the signs
$\beta_1$, $\beta_2$ and $\beta_3$, (\ref{eq:foe}) constitutes
eight systems of ordinary differential equations. Nevertheless,
this set of systems is easier to solve than second-order
(Euler-Lagrange) equations. In order to obtain a complete kink
solution we have to join solutions from the first-order
differential equations with different choices of the signs
$(-1)^{\beta_i}$ in different intervals covering the real line.
The reason for this is that the first-order differential equations
inherit the information of the second-order equations defined
piecewise. Assuming that we search for continuous and
differentiable solutions, the sequence of signs $(-1)^{\beta_i}$
corresponding to the different pieces that constitutes a solution
is prescribed. In section 4 we shall illustrate this approach in
several cases. From (\ref{eq:e3}) it is readily seen that the
energy of a solitary wave, solution of (\ref{eq:foe}) with only
one piece, depends only on the topological charge of the
solution. In this case, it is said that the Bogomol'nyi bound is
saturated. However, if the orbit $\lambda$ is given by
$\lambda=\cup_{j=1}^J \lambda^j$, where $J$ is the number of
pieces of $\lambda$ and $\lambda^j$ stands for the $j^{\rm th}$
piece, we have:
\begin{eqnarray}\label{eq:energy}
{\cal E}[\lambda]&=&\sum_{\mbox{\footnotesize pieces of }
\lambda} \int d x \frac{d W(\lambda)}{d x}=\sum_{j} \int
\sum_{i=1}^3\frac{\partial
W^{\{\beta_i\}_j}}{\partial\lambda_i}\,d\lambda_i\nonumber\\
&=&\sum_{j}^J (W^{\{\beta_i\}_j}(\lambda^j_{\mbox{\footnotesize
final}})-W^{\{\beta_i\}_j}(\lambda^j_{\mbox{\footnotesize
initial}}))\  ,
\end{eqnarray}
where $\{\beta_i\}_j$ represents the values of the $\beta_i$
parameters for the $j^{\rm th}$ piece of the solution.

\subsection{Solutions via quadratures}

In order to solve system (\ref{eq:foe}), we rewrite it in the
form:
\begin{equation}\label{eq:e6}
\frac{d\lambda_i}{(-1)^{\beta_i}2\sqrt{1-\lambda_i}\prod_{j=1}^4
(\lambda_i-c_j)}=\frac{dx}{f_i(\lambda)}\quad,\quad i=1,2,3\  ,
\end{equation}
where we have defined
$c=(\bar{\alpha}^2,\bar{\sigma}_2^2,\bar{\sigma}_3^2,\bar{\sigma}_4^2)$
and $\bar{\sigma}_4^2=0$ for future convenience. The sum of these
equations gives
\begin{equation}\label{eq:1foe}
\sum_{i=1}^3\frac{d\lambda_i}{(-1)^{\beta_i}2\sqrt{1-\lambda_i}
\prod_{j=1}^4(\lambda_i-c_j)}=0
\end{equation}

Multiplying each side of (\ref{eq:e6}) by $\lambda_i$ and summing
over $i$, we obtain:
\begin{equation}\label{eq:2foe}
\sum_{i=1}^3\frac{\lambda_id\lambda_i}{(-1)^{\beta_i}2\sqrt{1-\lambda_i}
\prod_{j=1}^4(\lambda_i-c_j)}=0\  .
\end{equation}

Also, multiplying (\ref{eq:e6}) by $\lambda_i^2$ and summing again
over $i$ we reach the equation that establishes the dependence of
the kink components on $x$
\begin{equation}\label{eq:3foe}
\sum_{i=1}^3\frac{\lambda_i^2d\lambda_i}{(-1)^{\beta_i}2\sqrt{1-\lambda_i}
\prod_{j=1}^4(\lambda_i-c_j)}=dx\  .
\end{equation}

We shall now determine the kink orbits and the form factor by
invoking (\ref{eq:1foe}), (\ref{eq:2foe}), and (\ref{eq:3foe}).
Integration of the first two equations,
\begin{displaymath}
\sum_{i=1}^3\frac{(-1)^{\beta_i}}{2}\int\frac{d\lambda_i}{\sqrt{1-\lambda_i}\prod_{j=1}^4
(\lambda_i-c_j)}=\gamma_2
\end{displaymath}
\begin{displaymath}
\sum_{i=1}^3\frac{(-1)^{\beta_i}}{2}\int\frac{\lambda_id\lambda_i}
{\sqrt{1-\lambda_i}\prod_{j=1}^4(\lambda_i-c_j)}=\gamma_3
\end{displaymath}
leads us to the expression of the generic kink orbits:
\begin{equation}\label{eq:gen1}
e^{2\gamma_2}=\prod_{j=1}^4\textstyle\left|\frac{\sqrt{1-\lambda_1}-\sqrt{1-c_j}}
{\sqrt{1-\lambda_1}+\sqrt{1-c_j}}\right|^{\frac{(-1)^{\beta_1}}{F_j(c)}}\cdot
\displaystyle\prod_{j=1}^4\textstyle\left|\frac{\sqrt{1-\lambda_2}-\sqrt{1-c_j}}
{\sqrt{1-\lambda_2}+\sqrt{1-c_j}}\right|^{\frac{(-1)^{\beta_2}}{F_j(c)}}\cdot
\displaystyle\prod_{j=1}^4\textstyle\left|\frac{\sqrt{1-\lambda_3}-\sqrt{1-c_j}}
{\sqrt{1-\lambda_3}+\sqrt{1-c_j}}\right|^{\frac{(-1)^{\beta_3}}{F_j(c)}}\nonumber
\end{equation}
\begin{equation}\label{eq:gen2}
e^{2\gamma_3}=\prod_{j=1}^3\textstyle\left|\frac{\sqrt{1-\lambda_1}-\sqrt{1-c_j}}
{\sqrt{1-\lambda_1}+\sqrt{1-c_j}}\right|^{\frac{(-1)^{\beta_1}c_j}{F_j(c)}}\cdot
\displaystyle\prod_{j=1}^3\textstyle\left|\frac{\sqrt{1-\lambda_2}-\sqrt{1-c_j}}
{\sqrt{1-\lambda_2}+\sqrt{1-c_j}}\right|^{\frac{(-1)^{\beta_2}c_j}{F_j(c)}}\cdot
\displaystyle\prod_{j=1}^3\textstyle\left|\frac{\sqrt{1-\lambda_3}-\sqrt{1-c_j}}
{\sqrt{1-\lambda_3}+\sqrt{1-c_j}}\right|^{\frac{(-1)^{\beta_3}c_j}{F_j(c)}},
\end{equation}
where $F_j(c)=\sqrt{1-c_j}\prod_{l=1,l\neq j}^4(c_j-c_{\,l})$, and
$\gamma_2$ and $\gamma_3$ are arbitrary real constants that
specify a particular kink orbit.

The integration of (\ref{eq:3foe})
\begin{displaymath}
\sum_{i=1}^3\frac{(-1)^{\beta_i}}{2}\int\frac{\lambda_i^2d\lambda_i}
{\sqrt{1-\lambda_i}\prod_{j=1}^4 (\lambda_i-c_j)}=\gamma_1+x
\end{displaymath}
gives us the form factor of the kink:
\begin{equation}\label{eq:gen3}
e^{2(\gamma_1+x)}=\prod_{j=1}^3\textstyle\left|\frac{\sqrt{1-\lambda_1}-\sqrt{1-c_j}}
{\sqrt{1-\lambda_1}+\sqrt{1-c_j}}\right|^{\frac{(-1)^{\beta_1}c_j^2}{F_j(c)}}\cdot
\displaystyle\prod_{j=1}^3\textstyle\left|\frac{\sqrt{1-\lambda_2}-\sqrt{1-c_j}}
{\sqrt{1-\lambda_2}+\sqrt{1-c_j}}\right|^{\frac{(-1)^{\beta_2}c_j^2}{F_j(c)}}\cdot
\displaystyle\prod_{j=1}^3\textstyle\left|\frac{\sqrt{1-\lambda_3}-\sqrt{1-c_j}}
{\sqrt{1-\lambda_3}+\sqrt{1-c_j}}\right|^{\frac{(-1)^{\beta_3}c_j^2}{F_j(c)}},
\end{equation}
$\gamma_1$ being an integration constant associated with the
translational invariance of the system. Expressions
(\ref{eq:gen1}), (\ref{eq:gen2}) and (\ref{eq:gen3}) provide us
with the whole manifold of solitary waves.

\subsection{Frontiers and barriers. Basic kinks}

We shall now prove that the set of solitary waves is confined to
living in a bounded region of the internal space, which in fact
corresponds to a parallelepiped in the elliptic space. For the
sake of clarity, we shall restrict our study to the range
$\bar{\alpha}^2\in L$, where $L=\bigcup_{i=1}^3L_i$ is the set in
which the kink manifold is richest, see Section 2. This include
the regimes $E$2, $H$1, and $H$2. Squaring the first equation in
(\ref{eq:e6}), and defining the generalized momentum
$\pi_1=g_{11}(\lambda)\frac{d\lambda_1}{dx}$, we have:
\begin{equation}\label{eq:e7}
\frac{1}{2}\pi_1^2-\frac{\lambda_1^2(\lambda_1-\bar{\alpha}^2)^2}{8(1-\lambda_1)}=0.
\end{equation}

Equation (\ref{eq:e7}) can be regarded as that governing the
motion of a particle moving under the influence of the potential
function
\begin{displaymath}
{\cal U}(\lambda_1)=\left\{
\begin{array}{cl}
-\displaystyle\frac{\lambda_1^2\left( \lambda_1-\bar{\alpha}^2
\right) ^2}{8\left(1-\lambda_1\right)
}\quad &,\quad -\infty<\lambda_1<\bar{\sigma}_3^2\\
\\
\infty &,\quad \bar{\sigma}_3^2<\lambda_1<\infty
\end{array}
\right.
\end{displaymath}

The function has at least one minimum in $\lambda_1=0$ and a
second one in $\lambda_1=\bar{\alpha}^2$ if $\bar{\alpha}^2\in
L_1$. Furthermore, the function ${\cal U}(\lambda_1)$ goes to
$-\infty$ as $\lambda_1$ tends to $-\infty$. Thus the bounded
motion can only occur in the interval $[0,\bar{\sigma}_3^2]$.
This, combined with the boundary conditions, leads us to the
conclusion that the kink solutions lie in the parallelepiped
$\bar{P}_3(0)=[0,\bar{\sigma}_3^2]\times[\bar{\sigma}_3^2,\bar{\sigma}_2^2]\times[\bar{\sigma}_2^2,1]$.

There is still more information that can be extracted following
this procedure, owing to the appearance of a second minimum. Let
us first fix a value $\bar{\alpha}^2$ in $L$, and let us set
$\bar{\alpha}^2\in L_i$ for some $i$ that depends on
$\bar{\alpha}^2$. Squaring the $i^{\rm th}$ equation of the system
(\ref{eq:e6}) and defining the generalized momentum
$\pi_i=g_{ii}(\lambda)\frac{d\lambda_i}{dx}$, we arrive at a
similar one-dimensional dynamics:
\begin{equation}
\frac{1}{2}\pi_i^2-\frac{\lambda_i^2(\lambda_i-\bar{\alpha}^2)^2}{8(1-\lambda_i)}=0.
\end{equation}

Accordingly the corresponding potential function is now defined
by ${\cal U}(\lambda_1)$ if $i=1$ and
\begin{displaymath}
{\cal U}(\lambda_i)=\left\{
\begin{array}{cl}
-\displaystyle\frac{\lambda_i^2\left( \lambda_i-\bar{\alpha}^2
\right) ^2}{8\left(1-\lambda_i\right)
}\quad &,\quad \textrm{min}\{L_i\}<\lambda_i<\textrm{max}\{L_i\}\\
\\
\infty &,\quad \lambda_i\notin L_i
\end{array}
\right.
\end{displaymath}
for $i=2,3$. The minimum $\lambda_i=\bar{\alpha}^2$ now separates
the bounded motion of the one-dimensional system into two
intervals  - the $\lambda_i\in
L_i^-=[\textrm{min}\{L_i\},\bar{\alpha}^2]$ interval and the
$\lambda_i\in L_i^+=[\bar{\alpha}^2,\textrm{max}\{L_i\}]$
interval -, and into the trivial motion
$\lambda_i=L_i^0=\bar{\alpha}^2$. This, together with the
asymptotic conditions, leads us to conclude that, besides living
in $\bar{P}_3(0)$, the kink solutions lie entirely in the sets
\[
\bar{P}_3(0)^{-,0,+}=\{\lambda\in\bar{P}_3(0)\quad\textnormal{with}\quad\lambda_i\in
L_i^{-,0,+}\}.\newline
\]

This decomposition of the parallelepiped $\bar{P}_3(0)$ is, for
the case we shall study in detail, regime $H1$, as follows (see
Figure 2):
\begin{eqnarray}
\bar{P}_3(0)&=&\bar{P}_3(0)^-\cup \bar{P}_3(0)^0\cup
\bar{P}_3(0)^+=
[0,\bar{\sigma}_3^2]\times[\bar{\sigma}_3^2,\bar{\alpha}^2]
\times[\bar{\sigma}_2^2,1]\, \cup \nonumber\\ &&
\cup\,[0,\bar{\sigma}_3^2]\times
\{\bar{\alpha}^2\}\times[\bar{\sigma}_2^2,1]\,
\cup\,[0,\bar{\sigma}_3^2]
\times[\bar{\alpha}^2,\bar{\sigma}_2^2]\times[\bar{\sigma}_2^2,1]\
. \nonumber
\end{eqnarray}

The parallelepipeds $\bar{P}_3(0)^-$ and $\bar{P}_3(0)^+$ contain
families of solutions that depend on two and three parameters,
whereas the plane $\bar{P}_3(0)^0$ only contains two-parametric
solutions.

Thus, introduction of the factor $(\lambda_i-\bar{\alpha}^2)^2$
into the potential function $U(\lambda)$ leads us (within our
range of study) to a new confinement of kink solutions in the
parallelepiped $\bar{P}_3(0)$. The generic kink solutions divide
into two sectors and, in addition to this, a new kind of
two-parametric solutions arises: those satisfying
$\lambda_i=\bar{\alpha}^2$. Consequently, the kink manifold can
be decomposed as follows:
\begin{equation}\label{DescompLi}
{\cal C}={\cal C}_i^{-}\sqcup {\cal C}_i^{0}\sqcup {\cal C}_i^+\,
,
\end{equation}
where ${\cal C}_i^{-,0,+}$ represent the class of kink solutions
with $\lambda_i\leq\bar{\alpha}^2$, $\lambda_i=\bar{\alpha}^2$ and
$\lambda_i\geq\bar{\alpha}^2$ respectively.

\subsection{Stability}

In this sub-section we discuss how to determine the stability
properties of the kink solutions.  For the whole variety of kink
solutions in this system, it is not possible to solve
$\lambda_1,\lambda_2$ and $\lambda_3$ in terms of elementary
functions of $x$. Therefore, it is not possible to explicitly
write out the Hessian operator for any kink in the model and,
hence, the stability properties cannot be studied through analysis
of its spectrum.

To determine the stability of the solutions, we use instead the
arguments developed in Ref. \cite{Nos3} based on the Jacobi fields
along kink solutions. Although the treatment depicted in that
paper is for a deformed Sigma $O(3)$ model, the extension to this
model can be readily carried out. Following this procedure, a rule
establishing the stability (instability) of the solutions is
obtained: each solution crossing either the edge
$F_1F_3\equiv\{\bar{\sigma}_3^2,\bar{\sigma}_3^2,\lambda_3\}$ or
the edge
$AF_2\equiv\{\lambda_1,\bar{\sigma}_2^2,\bar{\sigma}_2^2\}$
becomes an unstable solution, since these two edges constitute
lines of conjugate points of each vacuum of the theory.

The key point is that the superpotential function is not
differentiable over either of these two edges and, consequently,
the energy of the kink (\ref{eq:energy}) is not a topological
quantity since it depends on the value of the superpotential at
the crossing points.

In what follows, and bearing this remark in mind, we shall only
mention the character of each of the kinks described.

\section{Description of the Kink manifold in the H1 regime}

The description of the kink manifold in the different regimes
arising in our model is a long and tedious task. We shall
therefore focus our attention on a particular example: the $H1$
regime. Nevertheless, this case will suffice to illustrate the
general features that also arise in other regimes of our model.
We shall now describe the behaviour of the kinks that arise in
the $H1$ regime of our model. We can find basic kinks, similar to
the solutions $TK1$ and $TK2$ in MSTB model, that are placed on
the edges of the characteristic parallelepiped in the elliptic
space (see figures 4,5 and 6). These solutions are the simplest
kinks in our model and they consist of a single lump, such that
they can be interpreted as an extended particle. We shall show
that the kink manifold includes other kink solutions involving
several lumps associated with the basic kinks.

We recall some remarkable points of the $H$1 regime from the
previous sections: The number of minima is three in the
``elliptic" space, and eight in the Cartesian one (see Figure 2):
$ {\cal M}_2=\left\{\phi^{v_1}=(\pm 1,0,0), \phi^{v_2}=(\pm
\alpha,0,0),\phi^{v_3}=(\pm \frac{\alpha}{\sigma_3},0,\pm
\frac{\bar{\sigma}_3
}{\sigma_3}\sqrt{\bar{\alpha}^2-\bar{\sigma}_3^2})\right\}$. The
ellipsoid $E\equiv
\phi_1^2+\frac{\phi_2^2}{\bar\sigma_2^2}+\frac{\phi_3^2}{\bar\sigma_3^2}=1$
(that is, $\lambda=(0,\lambda_2,\lambda_3)$), the one-sheet
hyperboloid $H\equiv
\frac{\phi_1^2}{1-\bar{\alpha}^2}+\frac{\phi_2^2}{\bar\sigma_2^2-\bar{\alpha}^2}
-\frac{\phi_3^2}{\bar{\alpha}^2-\bar\sigma_3^2}=1$ (or
$\lambda_2=\bar{\alpha}^2$), and the planes $\phi_{2,3}=0$ are
distinguished surfaces in the internal space. In 3.3, we have
proved that all the topological solutions are confined within the
above-mentioned ellipsoid $E$. From this point of view, these
surfaces play the role of separatrices among three-parameter
families of solutions, as proved above. These solutions are
associated with finite values of the integration constants,
$\gamma_i$. It is usual in the literature \cite{Nos3} to refer to
this class of solutions as generic solutions. On the other hand,
these surfaces also contain the trajectories of two-parameter
families of solitary waves, which correspond to asymptotic values
of the constants $\gamma_i$. Accordingly, they are called
non-generic solutions.

Finally, we describe the kink manifold in these cases. We can
distinguish: A, Non-generic, two-parametric families, and B,
Generic, three-parametric families of solitary waves:

\begin{itemize}
\item[{\bf A}] Two-Parametric families of solutions:

\begin{itemize}
\item[{\bf A1}] Solutions on the ellipsoid $E$.

The potential term $U_1$ vanishes on this surface. Accordingly,
the superpotential function is:
\begin{displaymath}
W^{(\beta_2,\beta_3)}(\lambda_2,\lambda_3)=\sum_{i=2}^3W^{\beta_i}_i(\lambda_i)=\frac{1}{15}
\sum_{i=2}^3(-1)^{\beta_i}P_2(\lambda_i)\sqrt{1-\lambda_i}\
\quad,\quad\beta_i=0,1.
\end{displaymath}
The orbit of these solutions is given by
\begin{displaymath}
e^{2\gamma_2}=\prod_{j=1}^3\left|\frac{\sqrt{1-\lambda_2}-\sqrt{1-c_j}}
{\sqrt{1-\lambda_2}+\sqrt{1-c_j}}\right|^{\frac{(-1)^{\beta_2}c_j}{F_j(c)}}
\prod_{j=1}^3\left|\frac{\sqrt{1-\lambda_3}-\sqrt{1-c_j}}
{\sqrt{1-\lambda_3}+\sqrt{1-c_j}}\right|^{\frac{(-1)^{\beta_3}c_j}{F_j(c)}},
\end{displaymath}
$\gamma_2$ being an arbitrary real constant. We have two kind of
solutions:
\begin{itemize}
\item[$i)$] $\,T^{v_1,v_3}_E\,$: Stable topological solutions
that connect the minima $v^1$ and $v^3$ after having crossed the
plane $\phi_1=0$.

\item[${ii)}$] $\,N^{v_3}_E\,$: Unstable non-topological solutions
that join the minimum $v^3$ with itself. The trajectory of these
solutions starts from $v_3$, reaches the plane $\phi_1=0$, and
-after crossing the umbilical point $A$- returns to the same point
$v^3$; see Figure 4.
\begin{center}
\begin{figure}[htbp]
\epsfig{file=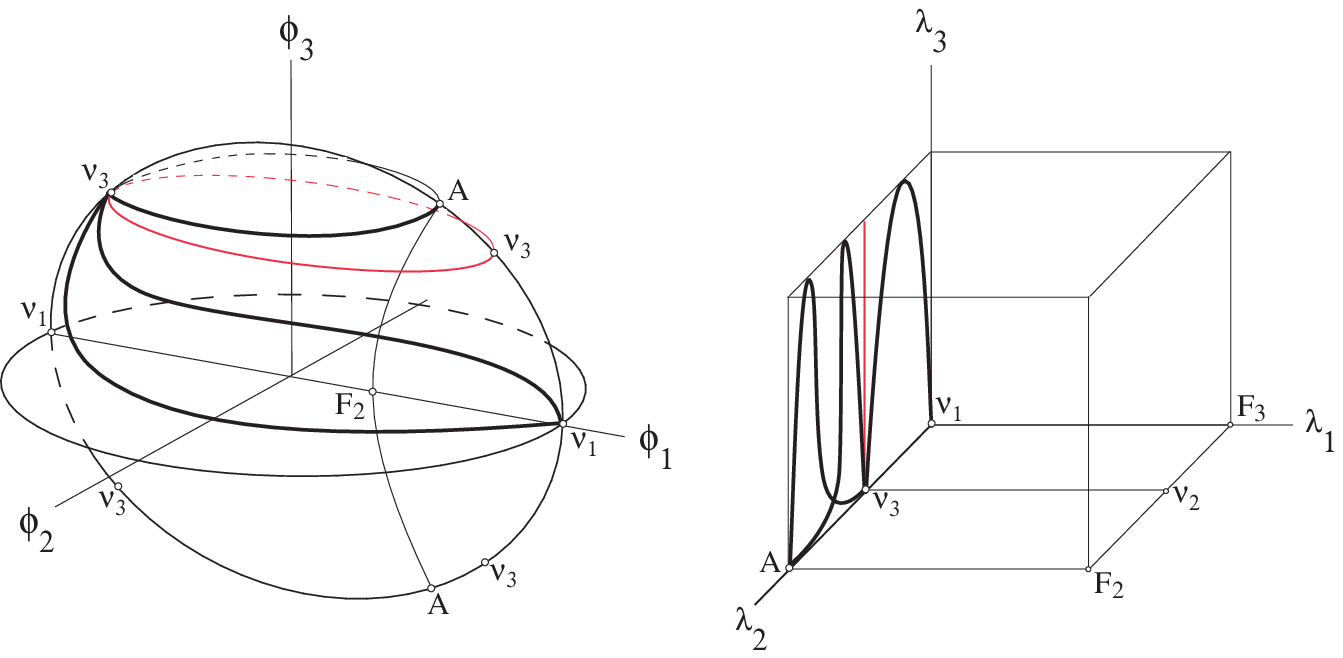, height=4.0cm} \epsfig{file=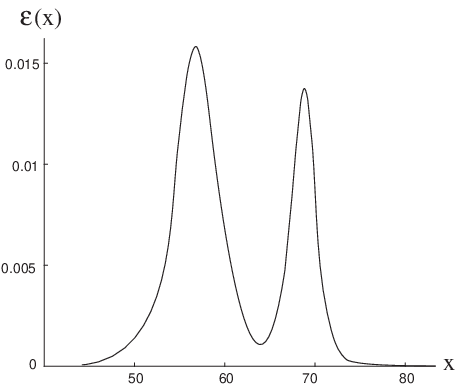,
height=4.0cm} \caption{\small \it Solitary waves on $E$ in the
Cartesian (left) and elliptic (middle) spaces. Energy density of
a kink of the family $T_E^{v_1,v_3}$ (right).}
\end{figure}
\end{center}
\end{itemize}
The energy of these solutions can easily be calculated by
integrating $dW$ along their respective orbits:
\begin{eqnarray}
E[T^{v_1,v_3}_{E}]&=&\int_{T^{v_1,v_3}_{E}}
dW=\int dW^{(0,0)}+\int dW^{(0,1)}\nonumber\\
&=&\int_{\bar{\sigma}_3^2}^{\bar{\alpha}^2}
dW^0_2+2\int_{\bar{\sigma}_2^2}^1dW^0_3=\frac{1}{15}\left(\alpha
P_2(\bar{\alpha}^2)
-\sigma_3P_2(\bar{\sigma}_3^2)-2\sigma_2P_2(\bar{\sigma}_2^2)\right)\nonumber\\
&=&\frac{2}{3}\left[\left(\frac{\alpha^5}{5}-\alpha^3\right)-\left(\frac
{\sigma_3^5}{5}-\sigma_3^3\right)-2\left(\frac{\sigma_2^5}{5}-\sigma_2^3\right)\right]\nonumber
\end{eqnarray}
\begin{eqnarray}
E[N^{v_3}_{E}]&=&\int_{N^{v_3}_{E}} dW=\int dW^{(0,0)}+\int
dW^{(0,1)}+\int dW^{(1,0)}+\int
dW^{(1,1)}\nonumber\\
&=&2\int_{\bar{\alpha}^2}^{\bar{\sigma}_2^2}dW_2^0+4\int_{\bar{\sigma}_2^2}^1dW_3^0
=\frac{1}{15}\left(-2\sigma_2P_2(\bar{\sigma}_2^2)-2\alpha
P_2(\bar{\alpha}^2) \right)\nonumber\\
&=&\frac{4}{3}\left[\left(\sigma_2^3-\frac{\sigma_2^5}{5}\right)+\left(\alpha^3-\frac{\alpha^5}{5}\right)
\right]\nonumber
\end{eqnarray}

%     \begin{figure}
%     \centerline{\includegraphics{Fig55.ps}}
%     \caption{Solitary waves}
%     \end{figure}

\item[{\bf A2}] Solutions on the plane $\phi_3=0$.

In this case, the terms $U_2$ and $U_1$ of the potential vanish,
but not simultaneously. The former vanishes over
$\lambda_2=\bar{\sigma}_3^2$, and the latter over
$\lambda_1=\bar{\sigma}_3^2$. Because of this, two superpotential
functions appear, and hence two systems of differential equations
must be involved in order to determine this solution.
Nevertheless, we can synthesize $W$ as follows:
\begin{displaymath}
W^{(\beta_k,\beta_3)}(\lambda_k,\lambda_3)=\frac{1}{15}\sum_{i=k,3}
(-1)^{\beta_i}P_2(\lambda_i)\sqrt{1-\lambda_i}\quad,\quad\beta_i=0,1,
\end{displaymath}
where $k=1$ for $\lambda_2=\bar{\sigma}_3^2$, and $k=2$ for
$\lambda_1=\bar{\sigma}_3^2$. The equations of the orbit on the
plane $\lambda_k=\bar{\sigma}_3^2$ are:
\begin{displaymath}
e^{2\gamma_2}=\prod_{j=1\atop{j\neq
3}}^4\left|\frac{\sqrt{1-\lambda_k}-\sqrt{1-c_j}}
{\sqrt{1-\lambda_k}+\sqrt{1-c_j}}\right|^{\frac{(-1)^{\beta_k}(c_j-c_3)}{F_j(c)}}
\prod_{j=1\atop{j\neq
3}}^4\left|\frac{\sqrt{1-\lambda_3}-\sqrt{1-c_j}}
{\sqrt{1-\lambda_3}+\sqrt{1-c_j}}\right|^{\frac{(-1)^{\beta_3}(c_j-c_3)}{F_j(c)}}.
\end{displaymath}
Again we have two kinds of solutions:
\begin{itemize}
\item[$i)$] $\,T^{v_1,v_2}_{\sigma_3}\,$: Unstable topological solutions linking the vacua $v^1$ and $v^2$. These solutions leave
$v^1$, intersect the axis $\phi_2$ and the segment $F_1F_3$
consecutively, and finally arrive at $v^2$, as depicted in Figure
5.

\item[$ii)$] $\,N^{v_2}_{\sigma_3}\,$: Unstable non-topological solutions connecting $v^2$. The solutions go from $v^2$, intersect
the axis $\phi_2=0$, cross the focus $F_2$, and return to the
initial point $v^2$.
\begin{center}
\begin{figure}[htbp] \epsfig{file=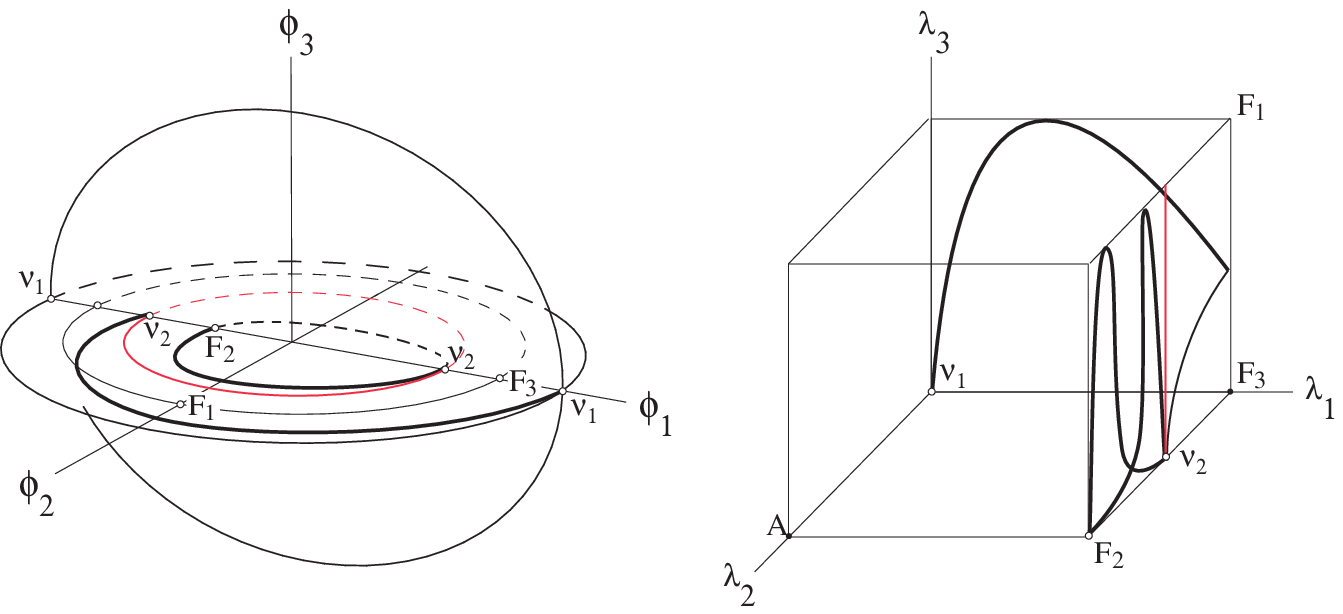,
height=4.0cm} \epsfig{file=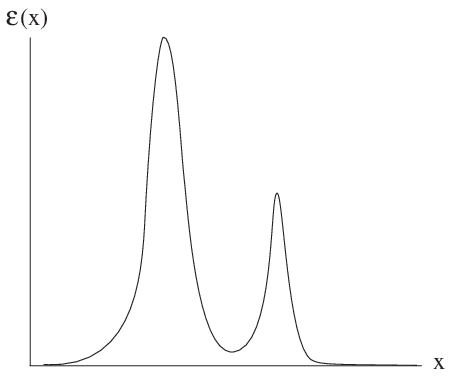, height=4.0cm}
\caption{\small \it Solitary waves on $\phi_3=0$ in the Cartesian
(left) and elliptic (middle) spaces. Energy density of a kink of
these families (right).}
\end{figure}
\end{center}

The computation of the energies is as follows:
\begin{eqnarray}
E[T^{v_1,v_2}_{\sigma_3}]&=&
\frac{2}{3}\left[\left(\frac{\alpha^5}{5}-\alpha^3\right)-\left(\frac{1}{5}-\alpha^2\right)
-2\left(\frac{\sigma_2^5}{5}-\sigma_2^3\right)\right]\nonumber\\
E[N^{v_2}_{\sigma_3}]&=&\frac{4}{3}\left[
\left(\alpha^3-\frac{\alpha^5}{5}\right)+\left(
\sigma_2^3-\frac{\sigma_2^5}{5}\right)\right]\, .\nonumber
\end{eqnarray}
\end{itemize}
\item[{\bf A3}] Solutions on the plane $\phi_2=0$, see Figure 6.

Now, the terms $U_3$ and $U_2$ of the potential vanish over
$\lambda_3=\bar{\sigma}_2^2$ and $\lambda_2=\bar{\sigma}_2^2$,
respectively. The two superpotential functions that appear can be
synthesized in a similar way:
\begin{displaymath}
W^{(\beta_1,\beta_k)}(\lambda_1,\lambda_k)=\frac{1}{15}\sum_{i=1,k}
(-1)^{\beta_i}P_2(\lambda_i)\sqrt{1-\lambda_i}\
\quad,\quad\beta_i=0,1,
\end{displaymath}
where $k=2$ for $\lambda_3=\bar{\sigma}_2^2$ and $k=3$ for
$\lambda_2=\bar{\sigma}_2^2$. The equations of the orbit on the
plane $\lambda_k=\bar{\sigma}_2^2$ are:
\begin{displaymath}
e^{2\gamma_2}=\prod_{j=1\atop{j\neq
2}}^4\left|\frac{\sqrt{1-\lambda_1}-\sqrt{1-c_j}}
{\sqrt{1-\lambda_1}+\sqrt{1-c_j}}\right|^{\frac{(-1)^{\beta_1}(c_j-c_2)}{F_j(c)}}
\prod_{j=1\atop{j\neq
2}}^4\left|\frac{\sqrt{1-\lambda_k}-\sqrt{1-c_j}}
{\sqrt{1-\lambda_k}+\sqrt{1-c_j}}\right|^{\frac{(-1)^{\beta_k}(c_j-c_2)}{F_j(c)}}\,.
\end{displaymath}
We now have three classes of solutions:
\begin{itemize}
\item[$i)$] $T^{v_1,v_2}_{\sigma_2}\,$: Stable topological solutions that join
the minima $v^1$ and $v^2$, as can be observed in Figure 6.

\item[$ii)$] $T^{v_3}_{\sigma_2}$: Unstable topological solutions
that connect the point $v^3$ with the minimum, which is its
reflection by the transformation $\phi_3 \rightarrow -\phi_3$,
previously crossing the focus $F_3$.

\item[$iii)$] $T^{v_2,v_3}_{\sigma_2}$: Unstable topological solutions that link the
points $v^2$ and $v^3$. In this case, the solutions depart from
$v^2$, and finally arrive at $v^3$ after intersecting the axis
$\phi_3$.

\begin{center}
\begin{figure}[htbp] \epsfig{file=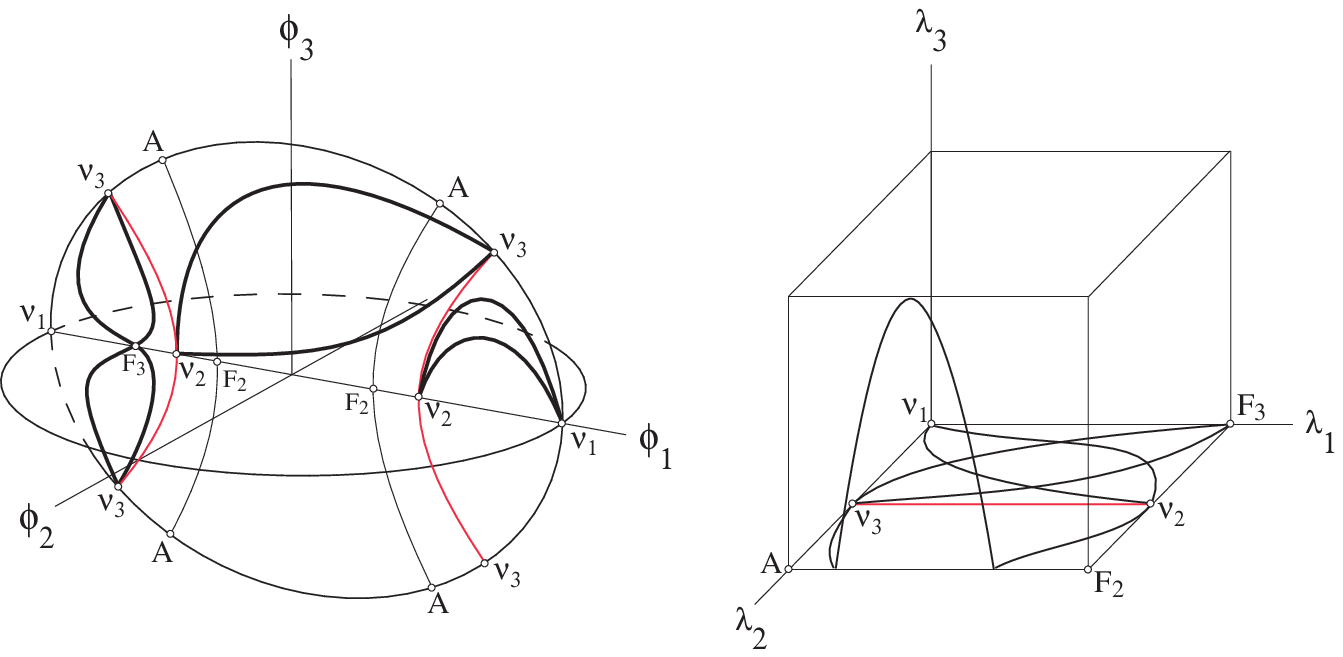,
height=4.0cm}\epsfig{file=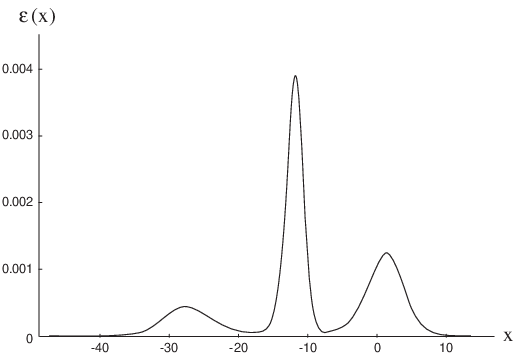, height=4.0cm}
\caption{\small \it Solitary waves on $\phi_2=0$ in the Cartesian
(left) and elliptic (middle) spaces. Energy density of a concrete
kink of the family $T_{\sigma_2}^{v_3}$ (right).}
\end{figure}
\end{center}

The energies for these solutions are:
\begin{eqnarray}
E[T^{v_1,v_2}_{\sigma_2}]&=&\frac{2}{3}\left[\left(
\frac{1}{5}-\alpha^2\right)-\left(\frac{\alpha^5}{5}-\alpha^3
\right)\right]\nonumber\\
E[T^{v_3}_{\sigma_2}]&=&\frac{4}{3}\left[\left(
\frac{1}{5}-\alpha^2\right)-\left(\frac{\alpha^5}{5}-\alpha^3
\right)\right],\nonumber
\end{eqnarray}
providing a simple kink energy sum rule:
$2E[T^{v_1,v_2}_{\sigma_2}]=E[T^{v_3}_{\sigma_2}]$. The remaining
energy is:
\begin{displaymath}
E[T^{v_2,v_3}_{\sigma_2}]=\frac{2}{3}\left[
\left(\frac{\sigma_3^5}{5}-\sigma_3^3\right)-2\left(\frac{\alpha^5}{5}-
\alpha^3\right)-\left(\frac{1}{5}-\alpha^2\right)\right].
\end{displaymath}
\end{itemize}

In figure 6(right), we have depicted the energy density
$\varepsilon(x)$ of a member of the family $T_{\sigma_2}^{v_3}$.
We notice that the kinks of this family consist of three basic
lumps.

\item[{\bf A4}] Solutions on the hyperboloid.

The term $U_2$ vanishes over $\lambda_2=\bar{\alpha}^2$ and hence
the superpotential function is:
\begin{displaymath}
W^{(\beta_1,\beta_3)}(\lambda_1,\lambda_3)=\frac{1}{15}\sum_{i=1,3}
(-1)^{\beta_i}P_2(\lambda_i)\sqrt{1-\lambda_i}\
\quad,\quad\beta_i=0,1.
\end{displaymath}
The equation of the orbit is:
\begin{displaymath}
e^{2\gamma_2}=\prod_{j=2}^4\left|\frac{\sqrt{1-\lambda_1}-\sqrt{1-c_j}}
{\sqrt{1-\lambda_1}+\sqrt{1-c_j}}\right|^{\frac{(-1)^{\beta_1}(c_j-c_1)}{F_j(c)}}
\prod_{j=2}^4\left|\frac{\sqrt{1-\lambda_3}-\sqrt{1-c_j}}
{\sqrt{1-\lambda_3}+\sqrt{1-c_j}}\right|^{\frac{(-1)^{\beta_3}(c_j-c_1)}{F_j(c)}}.
\end{displaymath}
In this case, only one family is found.
\begin{itemize}
\item $T^{v_2,v_3}_H\,$: The trajectories of these stable
solutions connect the points $v^2$ and $v^3$, previously
intersecting the plane $\phi_1=0$, as is shown in Figure 7.
Notice that the energy density in this case comprises two basic
lumps.
\end{itemize}
The energy is:
\[
E[T^{v_2,v_3}_H]=
\frac{2}{3}\left[\left(\frac{\sigma_3^5}{5}-\sigma_3^3\right)-\left(\frac{1}{5}-
\alpha^2\right)-2\left(\frac{\sigma_2^5}{5}-\sigma_2^3\right)\right]\,.
\]
\begin{center}
\begin{figure}[htbp]  \epsfig{file=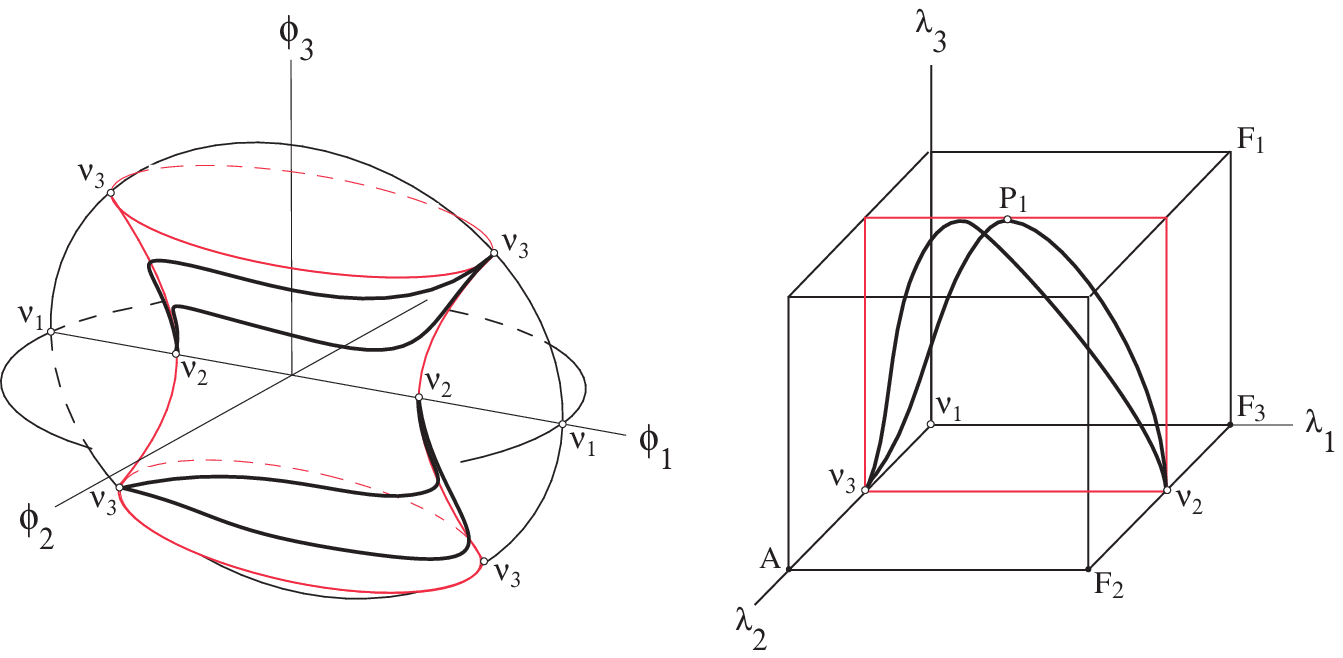, height=4.0cm}
\epsfig{file=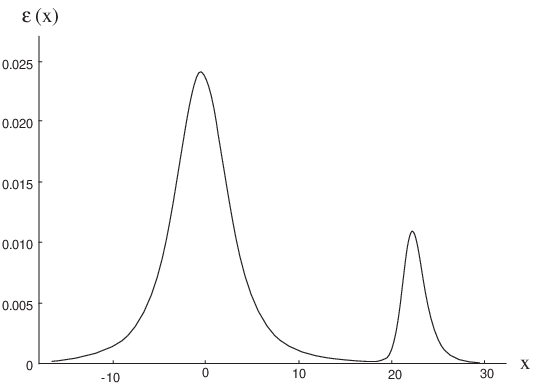, height=4.0cm}\caption{\small \it
Solitary waves on $H$ in the Cartesian (left) and elliptic
(middle) spaces. Energy density of a kink of this family (right).}
\end{figure}
\end{center}
\end{itemize}
\item[{\bf B}] Three-Parametric families of solutions. We find three kinds
of solutions:
\begin{itemize}
\item[{\bf B1}] Solutions located inside the ellipsoid and outside
the hyperboloid, see Figure 8:
\begin{itemize}
\item[$i)$] $T^{v_1,v_2}\,$: Stable topological solutions that join
$v^1$ and $v^2$. The solutions emerge from $v^1$, later cross the
plane $\phi_1=0$, and finally arrive at $v^2$.

\item[$ii)$] $T^{v_3}\,$: Unstable topological solutions, which
start from a minimum $v^3$, consecutively cross the planes
$\phi_1=0$ and $\phi_3=0$, intersecting the $F_1F_3$ edge, and
finally arrive at $v^3$. Notice that the energy density in this
case comprises four basic lumps.
\begin{center}
\begin{figure}[htbp]  \epsfig{file=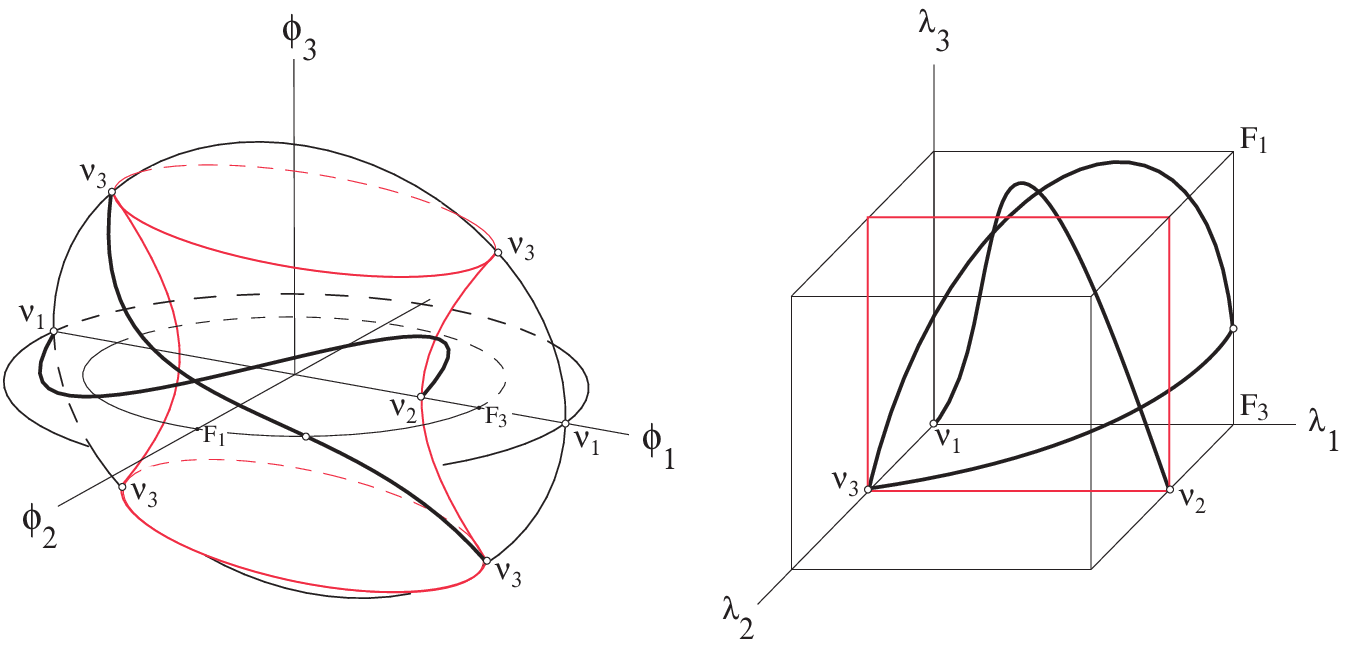,
height=4.0cm}\epsfig{file=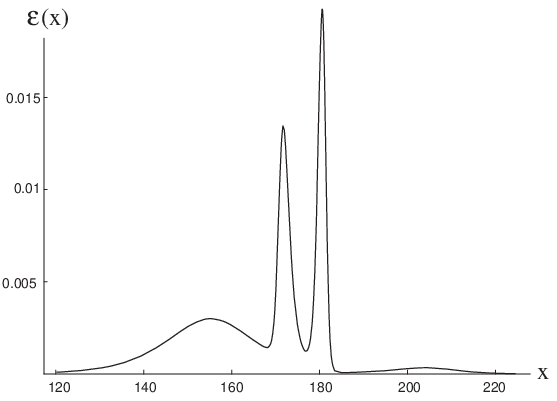, height=4.0cm}
\caption{\small \it Generic solitary waves in the Cartesian
(left) and elliptic (middle) spaces. Energy density of a kink of
the family $T^{v_3}$ (right).}
\end{figure}
\end{center}
Their energies are: \begin{eqnarray}
E[T^{v_1,v_2}]&=&\frac{2}{3}\left[\left(
\frac{\alpha^5}{5}-\alpha^3\right)-\left(\frac{1}{5}
-\alpha^2\right)-2\left(\frac{\sigma_2^5}{5}-\sigma_2^3\right)\right]\nonumber\\
E[T^{v_3}]&=&\frac{4}{3}\left[\left(\frac{\alpha^5}{5}-\alpha^3\right)
-\left(\frac{1}{5}-\alpha^2\right)-\left(\frac{\sigma_2^5}{5}-\sigma_2^3\right)\right]\nonumber
\end{eqnarray}
\end{itemize}
\item[{\bf B2}] Solutions located inside the hyperboloid:
\begin{itemize}
\item[$i)$] $T^{v_2,v_3}\,$: These are unstable solutions. They
leave $v^2$, cross the plane $\phi_1=0$, later intersect the
hyperbola $AF_2$, cross the plane $\phi_1=0$ again, and finally
arrive at the point $v^3$; see Figure 9.\end{itemize}
\begin{figure}[htbp] \hspace{4.5cm} \epsfig{file=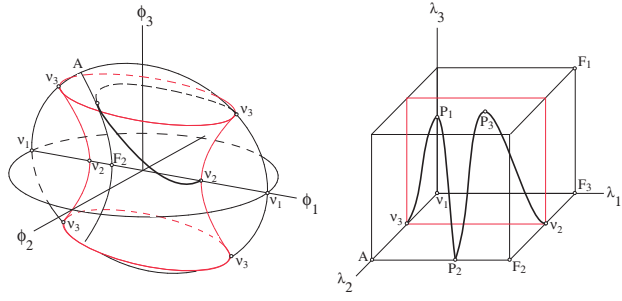,
height=4.0cm} \caption{\small \it Generic solitary waves in the
Cartesian (left) and elliptic (right) spaces.}
\end{figure}
The energy in this case is:
\[
E[T^{v_2,v_3}]=\frac{2}{3}\left[\left(\frac
{\sigma_3^5}{5}-\sigma_3^3\right)-\left(\frac{1}{5}-\alpha^2\right)-2\left(\frac
{\sigma_2^5}{5}-\sigma_2^3\right)-2\left(\frac{\alpha^5}{5}-\alpha^3\right)\right]
\]
\end{itemize}
To complete the previous energy calculations, the kink energy sum
rules satisfied by the generic solutions are offered:
\begin{itemize}
\item $E[T^{v_1,v_2}]=E[T^{v_1,v_2}_{\sigma_3}]$ \item
$2E[T^{v_2,v_3}]=E[N^{v_3}_E]+E[T^{v_2,v_3}_{H}]+E[T^{v_2,v_3}_{\sigma_2}]$\item
$2E[T^{v_3}]=E[T^{v_1,v_3}_E]+E[T^{v_2,v_3}_H]-3E[T^{v_1,v_2}_{\sigma_2}]$.
\end{itemize}

See Sub-section 3.1 of Reference \cite{Nos3} for an explanation
of the origin of these rules in a simpler setting. We stress that
the decomposition of the kink energy density in several lumps is
due to the kink energy sum rules.

\end{itemize}
Finally, as an example we depict the kink form factor (Fig. 10
and Fig. 11) for the two unstable generic solutions.
\begin{figure}[htbp] \hspace{2.30cm} \epsfig{file=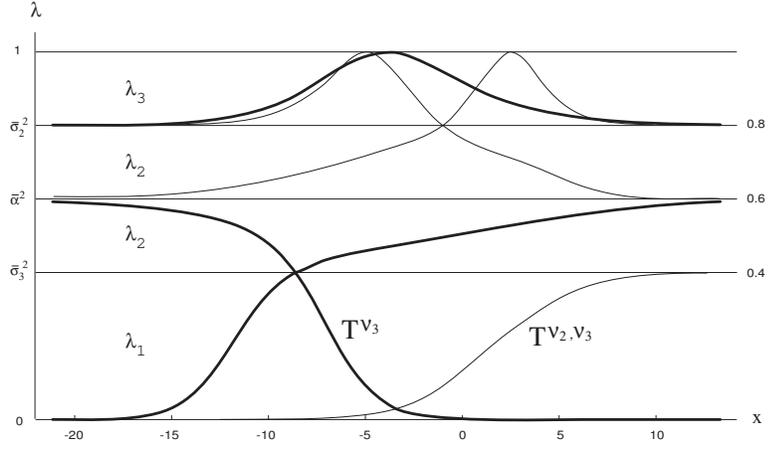,
height=6.0cm} \caption{\small \it Factor form for the
$T^{v_2,v_3}$ and $T^{v_3}$ solutions. For the $T^{v_3}$ solution,
we have taken $\gamma_1=0$, $\gamma_2=5$ and $\gamma_3=-5$,
whereas for the $T^{v_2,v_3}$ solution the constants are
$\gamma_1=\gamma_2=\gamma_3=0$.}
\end{figure}
\begin{figure}[htbp] \hspace{2cm} \epsfig{file=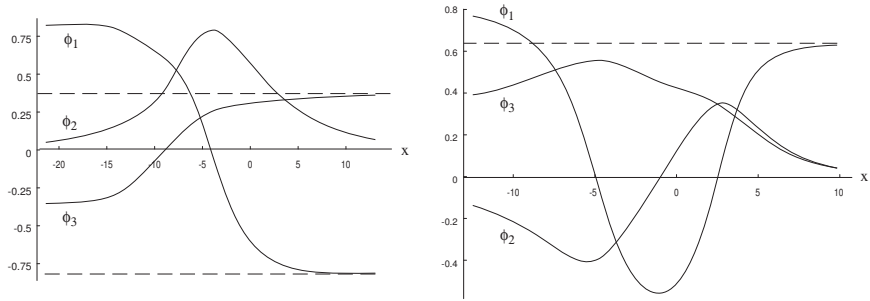,
height=4.0cm} \caption{\small \it Form factors in the Cartesian
space for the $T^{v_3}$ and $T^{v_2,v_3}$ solutions}
\end{figure}

\section{Further Comments}

It is possible to generalize this kind of model in two senses; we
enlarge the internal space with $N$ scalar fields and we include a
greater number of coupling constants $\bar{\alpha}_i^2$.

\noindent {\bf 1.} To study the generalization of this kind of
system to $N$ dimensions, it is first necessary to introduce
$N$-dimensional Jacobi elliptic coordinates. An appropriate
explanation of these can be seen in \cite{Nos3}. The potential
function we propose for the system is as follows:
\begin{displaymath}
U(\lambda;\bar{\alpha}^2)=\sum_{i=1}^NU_i(\lambda;\bar{\alpha}^2)=\frac{1}{2}
\sum_{i=1}^N\frac{\lambda_i^2
(\lambda_i-\bar{\alpha}^2)^2\prod_{j=2}^{N}(\lambda_i-\bar{\sigma}_j^2)}
{f_i(\lambda)}\,,
\end{displaymath}
where the coupling constants together with the coordinates
satisfies the chain
\begin{displaymath}
-\infty<\lambda_1<\bar{\sigma}_N^2<\lambda_2<\ldots<\lambda_{N-1}
<\bar{\sigma}_2^2<\lambda_N<1=\bar{\sigma}_1^2.
\end{displaymath}

The denominator is $f_i(\lambda)=\prod_{j\neq i}^N
(\lambda_i-\lambda_j)$, and $\bar{\alpha}^2$ is a real positive
constant. The function $U(\lambda;\bar{\alpha}^2)$ is positive
semi-definite and presents a number of zeroes, depending on
$\bar{\alpha}^2$. The most interesting kink manifold appears when
$\bar{\alpha}^2\in L_i\,,\,i=1,\ldots,N$, and becomes richer as
$i$ increases, the $L_i$ intervals being the trivial
generalization of those appearing in the three-dimensional
potential.

We shall now briefly study the vacuum manifold in all the $N$
different cases at once. Let us set $\bar{\alpha}^2$ such that
$\bar{\alpha}^2\in L_j$ for some $j$ between $1$ and $N$. To find
a zero of the function $U(\lambda;\bar{\alpha}^2)$, we must make
every term $U_i(\lambda;\bar{\alpha}^2)$ vanish. To visualize the
process, we shall seek help from the following graphic
\begin{center}
\includegraphics{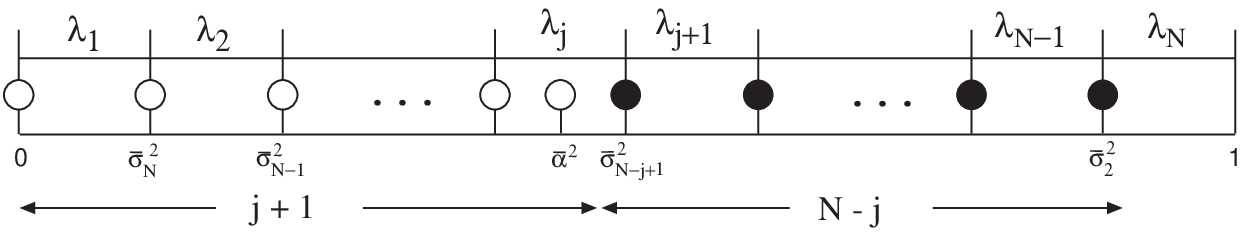}
\end{center}
Each circle in the $\lambda_k$ block represents a value that
$\lambda_k$ can take to make the term
$U_k(\lambda;\bar{\alpha}^2)$ null. Each value appearing in the
vacuum coordinates will be represented by a full circle, and
hence each vacuum in the elliptic space is represented by $N$
full circles. To fill the $N-j$ circles to the right of
$\bar{\alpha}^2$, there is only one possibility, as seen in the
figure, but to fill the remaining $j$ circles we have a number of
different ways equal to the number of permutations of $j+1$
elements, $j$ of them being repeated. Therefore, we have
$P^{(j+1)}_{j,1}=j+1$ zeroes of the $U(\lambda;\bar{\alpha}^2)$
function, $j$ of them being on the plane
$\lambda_j=\bar{\alpha}^2$. To figure out the number of
corresponding Cartesian vacua, we only need to take into account
the multiplicity of each elliptic vacuum. By doing this, we
conclude that by introducing a regular plane
$\lambda_j=\bar{\alpha}^2$ there are ${\cal V}=4+(j-1)2^{N-1}$
Cartesian vacua. The kink manifold thus decomposes into ${\cal
V}^{\,2}$ disconnected sectors \cite{AMJ1}.

\medskip

\noindent {\bf 2.} The second generalization considers not only
one parameter, $\bar{\alpha}^2$, but several of them. The
generalized potential is constructed as follows.

Let us consider numbers $n_i=0,1,2,\ldots$ with $i=1,\ldots,N$,
and let us define $(n_1+\ldots+n_N)$ different parameters
$\bar{\alpha}^2_{ij}$, such that for each $n_i\neq0$,
$\bar{\alpha}^2_{ij}\in L_i$ and $j=1,\ldots,n_i$. We can
therefore construct the $N$-dimensional potential:
\begin{equation}
U=\sum_{i=1}^N
U_i(\lambda,\bar{\alpha}_{ij}^2)=\frac{1}{2}\sum_{i=1}^N\frac{\lambda_i^2(\lambda_i-\bar{\sigma}_2^2)
(\lambda_i-\bar{\sigma}_3^2)}{f_i(\lambda)}\prod_{j=1\atop{n_i\neq0}}^{n_i}
(\lambda_i-\bar{\alpha}^2_{ij})^2\,.\label{genera}
\end{equation}

The case in which $\sum_{i=1}^Nn_i=0$ corresponds to the deformed
$O(N)$ linear sigma model \cite{Nos3} and the case
$\sum_{i=1}^Nn_i=1$, with $N=3$, is precisely the model studied in
the previous sections.

As $\sum_{i=1}^Nn_i$ increases, the vacuum manifold becomes more
and more abundant owing to the appearance of an increasing number
of roots in the potential. An easy way to account for the vacuum
manifold ${\cal V}$ is through the corresponding generalization
of the previous graphic
\begin{center}
\includegraphics{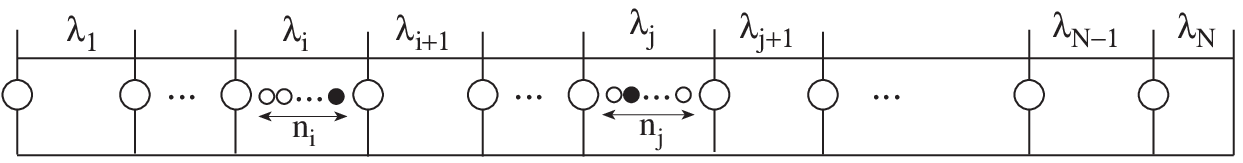}
\end{center}

In this picture $(n_1+\dots+n_N)$ additional circles appear,
$\alpha$-holes for short, since every $\bar{\alpha}_{ij}^2$ is
easily seen to be a root of the $U_i$ term in (\ref{genera}).
Computation of the number of vacua now proves to be an easy task
given that, as before, each vacuum point is represented by $N$
filled circles. It happens that the number of vacua -including
$v_1$, which corresponds to all the $\alpha$-holes emptied- is
given by:
\[
\textnormal{Card}({\cal V})=1+\sum_{q=1}^NN_q\,,
\]
where $N_q$ is the number of vacuum points with $q$ filled
$\alpha$-holes, which can be calculated readily using
combinatorial techniques.

Regarding the kink manifold, and looking at the corresponding
first-order equations, for each $\bar{\alpha}^2_{ij}$ we can
deduce a confinement of the solutions in $\bar{P}_3(0)$ similar to
that obtained in section 3. Therefore, a number of
$2^{(n_1+\ldots+n_N)}$ subsets of $\bar{P}_3(0)$ that host general
kink solutions appear.

The purpose of this construction is now clear. Recalling the
stability criterion and the confinement of the solutions due to
the factors $(\lambda_i-\bar{\alpha}^2_{ij})^2$, we can isolate
the edges
$F_1F_3=\{\bar{\sigma}_3^2,\bar{\sigma}_3^2,\lambda_3\}$ and
$AF_2=\{\lambda_1,\bar{\sigma}_2^2,\bar{\sigma}_2^2\}$. Proceeding
in this way, we can find subsets of the configuration space in
which only stable solutions emerge.

\subsection*{Acknowledgements}

The authors warmly acknowledge many discussions with J. Mateos
Guilarte. A.A.I. thanks the Secretaria de Estado de Educaci\'on y
Universidades of Spain for financial support.

\begin{thebiblio}{03}

{\small

\bibitem{Rajaraman} R. Rajaraman, {\sl Solitons and instantons. An introduction to solitons and instantons
in quantum field theory,} North-Holland Publishing Co., Amsterdam,
1982

\bibitem{Montonen} C. Montonen, {\sl Nucl. Phys.} {\bf B112} (1976)
349--357

S. Sarker, S.E. Trullinger and A.R. Bishop, {\sl Phys. Lett.}
{\bf A59} (1976) 255-258

K.R. Subbaswamy and S.E. Trullinger, {\sl Physica} {\bf D2}
(1981) 379-388

\bibitem{MT} E. Magyari and H. Thomas, {\sl Phys. Lett.} {\bf A100} (1984)
11--14

\bibitem{Ito} H. Ito, {\sl Phys. Lett.} {\bf A112} (1985) 119-123

\bibitem{ItoTasaki} H. Ito and H. Tasaki, {\sl Phys. Lett.} {\bf A113} (1985)
179-182

\bibitem{JMG}  J. Mateos Guilarte, {\sl Lett. Math. Phys.} 14 (1987) 169--176

J. Mateos Guilarte, {\sl Ann. Phys.} {\bf 188} (1988) 307--346

\bibitem{AMJ1} A. Alonso Izquierdo, M.A. Gonz\'alez Le\'on and J. Mateos
Guilarte, {\sl J. Phys.} {\bf A31} (1998) 209-229

\bibitem{Per} A. Perelomov, {\sl \lq\lq Integrable Systems of Classical
Mechanics and Lie Algebras"}, Birkh\"auser, Boston MA., 1990

\bibitem{Gar} R. Garnier, {\sl Ren. Circ. Mat. Palermo} 43 (1919)
155--191

\bibitem{Bazeia} D. Bazeia, J.R.S. Nascimento, R.F. Ribeiro and D. Toledo, {\sl J. Phys.}
 {\bf A30} (1997) 8157-8166

D. Bazeia, H. Boschi-Filho and F.A. Brito, {\sl JHEP} {\bf 9904}
(1999) 028

\bibitem{AMJ2} A. Alonso Izquierdo, M.A. Gonz\'alez Le\'on and J. Mateos Guilarte,
{\sl Phys. Rev.} {\bf D65} (2002) 085012

\bibitem{SV} M. Shifman and M. Voloshin, {\sl Phys. Rev.} {\bf
D 57} (1998) 2590

\bibitem{AMJM} A. Alonso Izquierdo, M.A. Gonz\'alez Le\'on, J. Mateos
Guilarte and M. de la Torre Mayado, {\sl Phys. Rev.} {\bf D66}
(2002) 105022

\bibitem{Va} S. Cecotti, C. Vafa, {\sl Comm. Math. Phys.} {\bf 158} (1993) 569--644

P. Fendley, S.D. Mathur, C. Vafa, and N.P. Warner, {\sl Phys.
Lett.} {\bf B243} (1990) 257--264

\bibitem{AMJ3} A. Alonso Izquierdo, M.A. Gonz\'alez Le\'on and J. Mateos
Guilarte, {\sl Phys. Lett.} {\bf B480} (2000) 373-380

\bibitem{Nos3} A. Alonso Izquierdo, M.A. Gonz\'alez Le\'on and J. Mateos
Guilarte, {\sl Nonlinearity}, {\bf 13} (2000) 1137-1169

\bibitem{Nos4} A. Alonso Izquierdo, M.A. Gonz\'alez Le\'on and J. Mateos Guilarte,
{\sl Nonlinearity}, {\bf 15} (2002) 1097-1125

\bibitem{NosAOP} A. Alonso Izquierdo, M.A. Gonz\'alez Le\'on and J. Mateos
Guilarte and M. de la Torre Mayado, {\sl Ann. Phys.} {\bf 308}
(2003) 664-691

}
\end{thebiblio}
\end{document}